\documentclass{emulateapj}
\slugcomment{to be appeared in ApJ}
\def\bmath#1{\mbox{\boldmath $#1$}}
\shorttitle{Relativistic Radiation MHD}
\begin{document}
\title{Numerical Treatment of Anisotropic Radiation Field Coupling
with the Relativistic Resistive Magnetofluids}
\author{Hiroyuki R. Takahashi\altaffilmark{1} and 
Ken Ohsuga\altaffilmark{2,3}}
\altaffiltext{1}{Center for Computational Astrophysics, National
  Astronomical Observatory of Japan, Mitaka, Tokyo 181-8588, Japan}
\altaffiltext{2}{Division of Theoretical Astronomy, National
  Astronomical Observatory of Japan, Mitaka, Tokyo 181-8588, Japan}
\altaffiltext{3}{School of Physical Sciences,Graduate University of
Advanced Study (SOKENDAI), Shonan Village, Hayama, Kanagawa 240-0193, Japan}

\begin{abstract}
We develop a numerical scheme for solving a fully special relativistic resistive
radiation magnetohydrodynamics. 
Our code guarantees conservations of total mass, momentum and energy. 
Radiation energy density and radiation flux are consistently updated
using the M-1 closure method,
which can resolve an anisotropic radiation fields
in contrast to the Eddington approximation as well as 
the flux-limited diffusion approximation.
For the resistive part, we adopt a simple form of the Ohm's law.
The advection terms are explicitly solved 
with an approximate Riemann solver, mainly HLL scheme,
and HLLC and HLLD schemes for some tests.
The source terms, 
which describe the gas-radiation interaction 
and the magnetic energy dissipation, are implicitly integrated,
relaxing the Courant-Friedrichs-Lewy condition 
even in optically thick regime 
or a large magnetic Reynolds number regime.  
Although we need to invert $4\times 4$ 
(for gas-radiation interaction) and $3\times 3$ 
(for magnetic energy dissipation) matrices 
at each grid point for implicit integration, 
they are obtained analytically without preventing 
massive parallel computing.
We show that our code gives reasonable outcomes
in numerical tests for ideal magnetohydrodynamics,
propagating radiation, and radiation hydrodynamics.
We also applied our resistive code 
to the relativistic Petschek type magnetic reconnection,
revealing the reduction of the reconnection rate via 
the radiation drag.
\end{abstract}

\section{Introduction}\label{intro}
Radiation and/or magnetic fields, relativity, and resistivity
play crucial roles in a number of high-energy astrophysical phenomena,
such as black-hole accretion-disks, jets, disk winds,
pulsar winds, magnetar flares, core collapse supernovae, 
and gamma-ray bursts.
For example, 
the geometrically thick disk is supported by the radiation 
pressure, which dominates the total pressure,
in the case of near- or super-critical accretion rate.
The radiation force is thought to accelerate the matter,
producing jets or winds \citep{1978PhyS...17..185L, 1980AJ.....85..329I,
1989A&A...216..294I,1996PASJ...48..529T}.
In contrast, the radiation drag
reduces the velocity of the relativistic outflow.
The magnetic field lines enhanced in the inner part of the
accretion disks launch jets/outflows \citep{1982MNRAS.199..883B,1985PASJ...37..515U,1997ApJ...476..632K}.
The magnetorotational instability (MRI) is thought to be 
origin of the disk viscosity, 
by which the angular momentum is transported outward
\citep{1059JETP...36..995,
1960PNAS...46..253C, 1991ApJ...376..214B}.
The resistivity would cause conversion 
from the magnetic energy to the energy of the matter
through the magnetic reconnection.
Also, the resistivity might influence the evolution and/or
saturation of MRI in the disks \citep{2007MNRAS.378.1471L, 2007A&A...476.1123F,2009ApJ...707..833S,2012arXiv1210.6664F}.

A global structure of the accretion disks and outflows
is investigated by radiation hydrodynamics (RHD) simulations 
\citep{1987ApJ...323..634E, 1988ApJ...330..142E,2000PASJ...52L...5O,
2005ApJ...628..368O,2006ApJ...640..923O},
magnetohydrodynamics (MHD) simulations \citep{2006PASJ...58..193M,
2008bhad.book.....K,2009MNRAS.394L.126M, 2010ApJ...711...50T}, and
Radiation-MHD (RMHD) simulations
\citep{2009PASJ...61L...7O,2010PASJ...62L..43T,2011ApJ...736....2O}.
Especially,
\cite{2010PASJ...62L..43T} showed high-velocity jets,
which is magnetically collimated,
are powered by the radiation force.
Also RMHD simulations of local patch of the disk are performed
\citep{2009ApJ...704..781H, 2013ApJ...767..148J}.
Although such works were great successful,
they should extent to relativistic simulations.

Many approximate methods have been proposed to solve the radiation transfer,
since the computational cost for rigorous method 
is too expensive to perform.
In the flux-limited diffusion (FLD) approximation,
a zeroth moment equation of the radiation transfer 
equation is solved to update the radiation energy density. 
The radiation flux as well as the radiation stress tensor
is given based on the gradient of the radiation energy density.
The FLD is a quite useful technique and gives appropriate
radiation fields within the optically thick regime,
but it does not always give precise radiation fields 
in the regime where the optical depth is around unity or less 
\citep[see][]{2011ApJ...736....2O}.
In contrast to the FLD approximation,
both zeroth and first moment equations are solved
in the Eddington approximation.
However, this method is somewhat problematic for
anisotropic radiation fields,
since the Eddington tensor is evaluated 
by assuming the isotropic radiation fields.
Additionally, the speed of light is effectively reduced in this method.

Although the variable Eddington tensor method
proposed by \cite{1992ApJS...80..819S} is known to give better results,
it is so complex and expensive.
One of the reasonable method is so-called M-1 closure,
in which the Eddington tensor is obtained 
as a function of the radiation energy density and radiation flux
\citep{1978JQSRT..20..541M, 1984JQSRT..31..149L}.
The anisotropy of radiation fields is approximately 
taken into consideration, and 
the radiation propagates with speed of light in an optically thin medium.
The M-1 closure is adopted to non-relativistic radiation
hydrodynamic code \citep{2007A&A...464..429G}, and recently to 
general relativistic (GR) code \citep{2013MNRAS.429.3533S}.
Another truncated moment
formalism of radiation fields in optically thick and thin limits
was proposed by \cite{2011PThPh.125.1255S}.

Relativistic RMHD or RHD simulations were recently
initiated.
\cite{2008PhRvD..78b4023F} first proposed a numerical scheme of
GR-RMHD, 
in which the Eddington approximation is employed.
\cite{2011MNRAS.417.2899Z} 
adopted a general relativistic RMHD code to the Bondi-Hoyle accretion on to the black holes. 
However, in their works, the explicit integration method is employed
even for the gas-radiation interaction.
In the relativistic phenomena,
the dynamical timescale as well as the timescale, 
that the characteristic wave passes the system,
could be comparable to the light crossing time.
Thus, although the numerical timestep becomes slightly short
via the explicit treatment of the propagating radiation,
the computational cost does not increase so much.
In contrast, if the absorption/scattering opacity is so large,
the timescale of gas-radiation interaction could be 
much shorter than the other timescales,
making the computation to be time consuming.
In the non-relativistic RHD/RMHD simulations,
such a difficulty is avoided by that 
the gas-radiation interaction terms are implicitly solved.
We should employ such an implicit treatment in the relativistic code
\citep{2012MNRAS.426.1613R,2013MNRAS.429.3533S,2013ApJ...764..122T}.

For resistive simulations, the magnetic energy dissipation should be
implicitly solved to relax the Courant-Friedrichs-Lewy condition 
in the regime of a large magnetic Reynolds number.
Here note that including the resistivity is a lot more complicated in
the relativistic MHD than in the non-relativistic MHD,
since we have to solve four additional equations for
calculating the time evolution of the electric fields
and charge density. The numerical treatment of relativistic MHD
simulations with resistivity were developed by authors
\citep{2007MNRAS.382..995K,2006ApJ...647L.123W,2009MNRAS.394.1727P,2011ApJ...735..113T}. 
The relativistic resistive RMHD simulations are challenging task.

In the present paper, 
we propose an explicit-implicit scheme for solving special relativistic
RMHD (SR-RMHD) and special relativistic Resistive RMHD (SR-R2MHD)
equations.
Here, the radiation fields in the observer frame are used
and we solve zeroth and first moment equations with the M-1 method.
Since the M-1 closure is constructed for the radiation
energy momentum tensor to be covariant,
the Lorentz transformation for the radiation fields is unnecessary
in our procedure.
Our scheme ensures a conservation of total energy and momentum 
(matter, magnetic field, and radiation). 
An advection of magnetofluids and the radiation 
is explicitly solved, 
and the gas-radiation interaction as well as
the magnetic energy dissipation via the resistivity
is implicitly treated.
Note that, although we propose SR code in the present study, the 
extension to the GR version would be straightforward except for 
the M-1 closure. The procedure of the M-1 closure in GR code is shown 
in \cite{2011PThPh.125.1255S}.


%

This paper is organized as follows: 
In \S~\ref{Method}, we introduce argument equations for SR-RMHD and
SR-R2MHD.
The numerical scheme is explained in \S~\ref{Numerical},
and we show the results in \S~\ref{test}.
Finally, \S~\ref{summary} is devoted to summary.

\section{Basic Equations}\label{Method}
In the following, we take a light speed as unity and assume the
Minkowski flat space-time. The metric is described by
$\eta^{\mu\nu}=\mathrm{diag}(-1,1,1,1)$.
Greek indices range over $0, 1, 2, 3$ and Latin does over
$1, 2, 3$, where $0$ indicates the time component and $1, 2, 3$ do space
components. 

A set of equations for the fully special relativistic radiation electro-magnetohydrodynamics
consists of conservation of mass,
\begin{equation}
\partial_\nu (\rho u^\nu) = 0,\label{geq:mcons}
\end{equation}
conservation of energy,
\begin{eqnarray}
 \partial_t\left[\rho h \gamma^2 - p_g
	    + \frac{\bmath E^2+\bmath B^2}{2}\right]
 +\nabla \cdot \left[\rho h \gamma u + \bmath E\times\bmath B\right]
 = G^0,\nonumber \\
 \label{geq:energy}
\end{eqnarray}
conservation of momentum,
\begin{eqnarray}
&& \partial_t
  \left[\rho h \gamma \bmath u + (\bmath E \times\bmath B)\right]
\nonumber \\
&&
  +\nabla \cdot
 \left[\rho h \bmath u \bmath u + p_g \bmath \delta 
  +\left(-\bmath E \bmath E - \bmath B \bmath B 
    + \bmath \delta\frac{|\bmath E|^2 + |\bmath B|^2}{2}\right)
 \right]
\nonumber \\
&& = \bmath G,\label{geq:momentum}
\end{eqnarray}
Maxwell equations
\begin{eqnarray}
 \partial_t \bmath B+\nabla \times \bmath E = 0,\label{geq:faraday}\\
 \partial_t \bmath E-\nabla \times \bmath B =-\bmath j,\label{geq:ampere}
\end{eqnarray}
where $\bmath \delta$ is the Kronecker delta. $\rho, p_g$ and $h$ are
the proper mass density, gas pressure and gas specific enthalpy.  
The exchange of the energy and momentum between the gas and the radiation,
$G^0$ and $G^i$ are shown in equations (\ref{geq:G0})-(\ref{geq:Gi}).

The bulk four velocity $u^\mu$ is related to the three
velocity $\bmath v$ by
\begin{eqnarray}
 u^\mu = \gamma (1, \bmath v),\label{geq:4vel}
\end{eqnarray}
where $\gamma=\sqrt{1+|\bmath u|^2}$ is the Lorentz factor.

Electric $\bmath E$ and magnetic $\bmath B$ fields are redefined to
absorb a factor of $1/\sqrt{4\pi}$. 
We should specify Ohm's law to relate the charge density $\bmath j$ and
$\bmath E$. 
When we assume an ideal MHD, the closure relation is given by
\begin{equation}
 \bmath E = -\bmath v \times \bmath B. \label{geq:ideal}
\end{equation}
Then, electric fields are determined without solving equation
(\ref{geq:ampere}). 

For a resistive MHD, we adopt a simple form of the
Ohm's law:
\begin{equation}
 \bmath j = \rho_e \bmath v 
  + \eta^{-1}\gamma\left[\bmath E + \bmath v\times \bmath B 
		    - (\bmath v \cdot \bmath E)\bmath v\right],
 \label{geq:resistive}
\end{equation}
where $\eta$ is an electric resistivity
\citep{1993PhRvL..71.3481B}, and  $\rho_e$ is the charge density,
which is obtained by solving charge conservation equation,
\begin{equation}
 \partial_t \rho_e + \nabla\cdot \bmath j = 0,\label{geq:charge}
\end{equation}
\citep{2007MNRAS.382..995K}. 
Since $\bmath E$ should evolve according to equation (\ref{geq:ampere}),
we have to solve four additional equations in 
relativistic resistive MHD. In our numerical code, we can switch on/off
the resistivity. 

Equations (\ref{geq:faraday})-(\ref{geq:ampere})
satisfy divergence conditions $\nabla \cdot \bmath B = 0$, and $\nabla
\cdot \bmath E = \rho_e$, if they are satisfied at the initial
state. But these conditions are violated due to numerical
errors. We adopted a Generalized Lagrange Multiplier (GLM) method
\citep{2002JCoPh.175..645D, 2007MNRAS.382..995K} to overcome these
problems. We do not describe details of this scheme, but it appears in
their papers.

The radiation field obeys following conservation equation
\begin{equation}
 T^{\mu\nu}_{\mathrm{rad},\nu} = -G^\mathrm{\mu},\label{geq:Tradcons}
\end{equation}
where the energy momentum tensor of radiation
$T^{\mu\nu}_{\mathrm{rad}}$ is given by
\begin{equation}
 T_\mathrm{rad}
  =\left(\begin{array}{cc}
    E_r, & \bmath F_r \\
	  \bmath F_r, & \bmath P_r\label{geq:Trad}
	 \end{array}\right).
\end{equation}
where $E_r$, $\bmath F_r$ and $\bmath P_r$ are the radiation energy density,
flux and stress measured in the laboratory frame.

The radiation exchanges its energy and momentum with fluids by
absorption/emission and scattering processes through the radiation four
force $G^\mu$:
\begin{eqnarray}
 G^0 &=& -\rho \kappa 
  \left(4\pi \mathrm{B} \gamma - \gamma E_r 
+ u_j F_r^j \right)\nonumber \\
 &&- \rho \sigma_s
\left[\gamma |\bmath u|^2 E_r + \gamma u_j u_k P^{jk}_r
-\left(\gamma^2 + |\bmath u|^2\right) u_j F_r^j \right],\label{geq:G0}
\end{eqnarray}
and 
\begin{eqnarray}
 G^{j} &=&- 4\pi \rho \kappa \mathrm{B} u^j
 + \rho (\kappa + \sigma_s)(\gamma F_r^j-u_kP^{jk}_r)\nonumber \\
&& -\rho \sigma_s u^j
\left(\gamma^2 E_r - 2\gamma u_k F^k_r + u_k u_l P_r^{kl}\right),\label{geq:Gi}
\end{eqnarray}
where $\kappa$ and $\sigma_s$ are absorption and
scattering coefficients measured in the comoving frame
\citep[e.g,][]{2008bhad.book.....K}. 

The blackbody intensity $\mathrm{B}$ is described by gas
temperature $T_g$ by
 \begin{equation}
  \mathrm{B}= \frac{a_R T_g^4}{4\pi},\label{geq:bb}
 \end{equation}
where $a_R$ is the radiation constant. The gas temperature is
determined by the Boyle--Charle's law:
\begin{equation}
p_g =  \frac{\rho k_B T_g}{\mu m_p},\label{geq:eos}
\end{equation}
where $k_B$ and $m_p$ are the Boltzmann constant and proton mass,
and $\mu$ is a mean molecular weight.

Finally, closure relations should be provided
by specifying the equation of state for the matter and radiation fields.
For the fluids, we assume a constant $\Gamma$-law, relating the specific
enthalpy with the gas pressure by
\begin{equation}
 h = 1 + \frac{\Gamma}{\Gamma-1}\frac{p_g}{\rho}.\label{eq:enthalpy}
\end{equation}
where $\Gamma$ is a specific heat ratio.

For the radiation field, $P^{jk}_r$ is assumed to be related
to $E_r$ and $F^j_r$ through the Eddington tensor
$P_r^{jk}=D_r^{jk} E_r$. 
In this paper, we assume a M-1 closure given by
\cite{1984JQSRT..31..149L},
which is explicitly described as 
\begin{eqnarray}
 D_r^{jk}&=&\frac{1-\chi}{2}\delta^{jk} + \frac{3\chi-1}{2}n^j n^k,
  \label{geq:DtensM1}\\
  \chi &=& \frac{3 + 4 |\bmath f|^2}{5 + 2 \sqrt{4 - 3 |\bmath f|^2}},\\
 f^j &=& \frac{F_r^j}{E_r},\\
 n^j &=& \frac{F_r^j}{|\bmath F_r|}. \label{geq:DtensM1end}
\end{eqnarray}
We have to note that the Eddington tensor of the M-1 model
is a function of $E_r$ and $F^j_r$, which can be
evaluated in the laboratory frame. 
For the Eddington approximation, which is another class of the closure
relation, the Eddington tensor
$D_r'^{jk}=\delta^{jk}/3$ should be evaluated at the comoving frame.  
Then we need to perform the Lorentz transformation to obtain $P_r^{jk}$ 
from $D_r'^{jk}$, $E_r$, and $F_r$ \citep{2013ApJ...764..122T}.
On the other hand, the M-1 closure is constructed for the radiation
energy momentum tensor to be covariant. 
Thus we can directly obtain $P_r^{jk}$ from $E_r$ and $F_r^j$ without
Lorentz transformation.

Here we note that a M-1 closure given by \cite{1984JQSRT..31..149L} is useful 
in the non-relativistic or special relativistic cases. The extension to 
general relativity is proposed by \cite{2011PThPh.125.1255S}.

Now we have 12 hyperbolic equations for SR-RMHD and 16 hyperbolic
equations for SR-R2MHD. When GLM method is adopted to
preserve divergence free conditions, 13 and 18 
equations should be numerically solved for SR-RMHD and SR-R2MHD, respectively. 

\section{Numerical Scheme}\label{Numerical}
In this section, we show how to solve SR-RMHD and SR-R2MHD
equations.
First, we show a numerical scheme to solve
SR-RMHD equation in \S~\ref{numR2MHD}. 
Next we show how to extent SR-RMHD code to SR-R2MHD by taking into account an
electric resistivity in \S \ref{numR3MHD}. 

\subsection{SR-RMHD}\label{numR2MHD}
Summarizing, an argument system of SR-RMHD is
\begin{eqnarray}
&&\partial_t D + \nabla \cdot (D \bmath v) = 0,\label{eq:r2mcons}\\
&&\partial_t e + \nabla \cdot \bmath m = G^0,\label{eq:r2e}\\
&&\partial_t \bmath m + \nabla \cdot \bmath \Pi = \bmath G,\label{eq:r2m}\\
&&\partial_t \bmath B + \nabla \times \bmath E = 0,\label{eq:r2b}\\
&&\partial_t E_r + \nabla \cdot \bmath F_r = - G^0,\label{eq:r2er}\\
&&\partial_t \bmath F_r + \nabla \cdot \bmath P_r = - \bmath G\label{eq:r2fr}
\end{eqnarray}
where 
\begin{eqnarray}
&& \bmath E = -\bmath v \times \bmath B,\\
&& D = \rho \gamma, \\
&& e = \rho h \gamma^2 - p_g + \frac{|\bmath E|^2 + |\bmath B|^2}{2},\\
&& \bmath m = \rho h \gamma \bmath u + \bmath E \times \bmath B,\\
&& \bmath \Pi= \rho h \bmath u \bmath u -\bmath \delta p_g
  -\bmath E \bmath E -\bmath B\bmath B 
  + \frac{\bmath \delta}{2}\left(|\bmath E|^2 + |\bmath B|^2\right).
\end{eqnarray}
In the Cartesian coordinate, the system can be described by a simple
phase equation
\begin{equation}
 \frac{\partial \mathcal{U}(\mathcal{P})}{\partial t} + \frac{\partial
  \mathcal{F}^k(\mathcal{P})}{\partial x^k} = \mathcal{S}(\mathcal{P}),
\end{equation}
where $\mathcal{P}$, $\mathcal{U}$, $\mathcal{F}$, and $\mathcal{S}$ are
primitive variables, conserved variables, fluxes, and source terms, 
\begin{equation}
 \mathcal{P} = 
  \left(\begin{array}{c}
   \rho \\
   u^j \\
   p_g \\
   B^j \\
   E_r \\
   F_r^j\\
\end{array}\right),\ 
 \mathcal{U} 
  = \left(\begin{array}{c}
     D \\
     m^j\\
     e\\
     B^j\\
     E_r\\
     F_r^j\\
  \end{array}\right),\ 
  \mathcal{F}^k
  = \left(\begin{array}{c}  
     D v^k\\
     \Pi^{jk}\\
     m^k \\
     \varepsilon^{jkl} E_l\\
     F_r^k \\
     P_r^{jk}
   \end{array}\right),
\end{equation}
and 
\begin{equation}
 \mathcal{S} 
  \equiv \left(\begin{array}{c}
     0 \\
     -S_E\\
     -S_F^j\\
     0\\
     S_E\\
     S_F^j\\
  \end{array}\right) 
=
 \left(\begin{array}{c}
     0 \\
     G^0\\
     G^j\\
     0\\
     -G^0\\
     -G^j\\
  \end{array}\right),\label{def:source}
\end{equation}
where $\varepsilon^{jkl}$ is the Levi-Civita antisymmetric tensor.
In the following, we consider 1-dimensional problems along the
$x$-direction without a loss of generality. 
\begin{equation}
 \frac{\partial \mathcal{U}}{\partial t} 
+\frac{\partial \mathcal{F}^x}{\partial x} = \mathcal{S}.\label{eq:1dform}
\end{equation}
Extension to multidimensional problems and to curved space is straightforward. 

We solve equation (\ref{eq:1dform}) using operator-splitting method as:
\begin{eqnarray}
  \frac{\partial \mathcal{U}}{\partial t} + \frac{\partial
   \mathcal{F}}{\partial x}= 0, \label{eq:advection}\\
  \frac{\partial \mathcal{U}}{\partial t} = \mathcal{S}.\label{eq:source}
\end{eqnarray}
The  conservative discretization of 1-dimensional equations
(\ref{eq:advection}) and (\ref{eq:source}) over a time step $\Delta t$ from $t = n\Delta t$ is
\begin{eqnarray}
 \mathcal{U}^*_i = \mathcal{U}^n_i -
  \frac{\Delta t}{\Delta x}\left(f_{i+1/2}^{n} -
			    f_{i-1/2}^n\right),\label{eq:dadvection}\\
 \mathcal{U}^{n+1}_i = \mathcal{U}^*_i
  +\mathcal{S}_i^{n+1}\Delta t,\label{eq:dsource}
\end{eqnarray}
where $\Delta x$ is the grid spacing and $i$ denotes the grid point,
$x=i\Delta x$. $f$ is the numerical flux described below.

Here, the equation (\ref{eq:advection}) is integrated explicitly,
while equation (\ref{eq:source}) is solved implicitly
\citep{2012MNRAS.426.1613R,2013MNRAS.429.3533S,2013ApJ...764..122T}.
Although a absorption or scattering timescales, 
$\sim 1/(\rho \kappa c)$ or $\sim 1/(\rho \sigma_s c)$
can be much shorter than the dynamical time scale 
in an optically thick medium,
the implicit integration of equation (\ref{eq:source}) allows us to
take the time step, $\Delta t$, larger than absorption/scattering timescales.
Since equation (\ref{eq:advection}) has a hyperbolic form, 
$\Delta t$ in our code is determined using maximum
wave velocities for radiation field $\lambda_r$ and magnetofluids
$\lambda_f$ as $\Delta t=C_\mathrm{cfl}\Delta x/\mathrm{max}(|\lambda_r|,
|\lambda_f|)$, where $C_\mathrm{cfl}<1$ 
is a Courant-Friedrichs-Lewy (CFL) number and
$\lambda_r$ and $\lambda_f$ are obtained 
by computing maximum values of eigenvalues for radiation
fields and magnetofluids (discussed later).

For the 1st step, we compute surface values of primitive variables
$\mathcal{P}_{i\pm 1/2,s}$ from cell centered variables
$\mathcal{P}_{i}$ as
\begin{eqnarray}
 \mathcal{P}_{i\pm \frac{1}{2},s} = \mathcal{P}_i 
  \pm \frac{\delta_x \mathcal{P}}{2},\label{eq:phalf}
\end{eqnarray}
where $s=L (R)$ denotes left (right) state variables. 
The spatial accuracy of numerical codes depends on the choice of slope
$\delta_x \mathcal{P}$.
Many types of slope limiter which preserve monotonicity are proposed. 
In this paper, we utilize a harmonic mean proposed by
\cite{1977JCoPh..23..263V},
which is a second order accuracy in space;
\begin{equation}
 \delta_x \mathcal{P}= 
\frac{2 \mathrm{max(0,\Delta \mathcal{P}_+ \Delta \mathcal{P}_-)}}
{\Delta \mathcal{P}_+ + \Delta \mathcal{P}_-},\label{eq:dxp}
\end{equation}
where
\begin{equation}
 \Delta \mathcal{P}_\pm = \pm (\mathcal{P}_{i\pm1} - \mathcal{P}_i).
\label{eq:dp}
\end{equation}

Extension to higher order schemes 
are straightforward \citep[e.g.][]{1984JCoPh..54..174C, 1996JCoPh.123....1M,
1999MNRAS.308.1069K, 2002A&A...390.1177D}.

For the 2nd step, numerical fluxes $f_{i \pm 1/2}$ are computed from
reconstructed primitive variables $\mathcal{P}_{i\pm 1/2,s}$.
We adopt an approximate Riemann solver to evaluate numerical fluxes. 
We utilize the HLL 
\citep{1983siamRev...25..35..61} scheme
to evaluate $f_{i\pm 1/2}$ given by
\begin{equation}
f_{i\pm 1/2}=
 \frac{\lambda^+\mathcal{F}_{L} 
- \lambda^-\mathcal{F}_{R} 
  +\lambda^+\lambda^-
  (\mathcal{U}_{R} - \mathcal{U}_{L})}
  {\lambda^+ - \lambda^-}.\label{eq:hllf}
\end{equation}
where $\lambda^+$ and $\lambda^-$ are maximum and minimum wave velocity,
respectively. The wave velocity is obtained by computing eigenvalues of
Jacobian matrix $\partial \mathcal{F}/\partial U$.
We note that the wave speed of radiation fields is independent of
fluid quantities when we utilize the M-1 closure. In other words,
Jacobian matrix of radiation fields is only a function of $E_r$ and
$F^j_r$ since the Eddington tensor is only a function of
$E_r$ and $F^j_r$. This indicates that the Jacobian matrix can be
completely decomposed into submatrices for magnetofluids and the
radiation. We can compute eigenmodes of magnetofluid and radiation
independently. 
For the radiation field, wave velocities, $\lambda_r^\pm$, are numerically computed 
from the Jacobian matrix and tabulated before time integration in our scheme, 
since the computation is time consuming
\citep{2007A&A...464..429G}. Here, we note that 
such a wave velocity is overestimated when the system is highly optically thick. 
In this limit, the radiation energy should be slowly diffused out with the 
diffusion velocity, $c/\tau$, in the comoving frame, where $\tau$ is the optical 
thickness. However, the eigen value computed from the Jacobian matrix has a large 
value $c / \sqrt{3}$, causing a large numerical diffusion. Thus, following to 
\cite{2013MNRAS.429.3533S}, we modify the wave velocities as
\begin{eqnarray}
 \lambda'^+ &\rightarrow
  \mathrm{min}\left(\lambda'^+,\frac{4}{3\tau^i}\right),\\
 \lambda'^- &\rightarrow \mathrm{max}\left(\lambda'^-,-\frac{4}{3\tau^i}\right),
\end{eqnarray}
where $\lambda'^\pm$ are the right- and left- going wave velocities in
the comoving frame and $\tau^i$ is the total optical depth in a cell.
This modification drastically reduces numerical diffusion in the optically
thick case.

For the MHD, wave speeds are computed by solving
quartic equation
\begin{equation}
 \rho h (1-c_s^2)a^4 = (1-\lambda^2)\left[(|b|^2 + \rho h c_s^2)a^2 -
				     c_s^2 \mathcal{B}^2\right],
\end{equation}
with $a=\gamma (\lambda - v^x)$ and $\mathcal{B}=b^x - \lambda
b^0$ \citep{2006MNRAS.368.1040M}. Here $c_s$ is the sound speed and
$b^\mu=\gamma(\bmath B\cdot
\bmath v, \bmath B/\gamma^2 + \bmath v(\bmath v \cdot \bmath B))$ is the
covariant form of the magnetic fields. 
The fast magnetosonic wave velocities $\lambda_f^{\pm}$ are obtained by
taking the maximum and minimum values of roots $\lambda$.
Numerical fluxes for magnetofluids are computed from equation
(\ref{eq:hllf}) using $\lambda_f^{\pm}$.

We note that higher order approximate Riemann solvers such as 
HLLC \citep{2006MNRAS.368.1040M, 2007JCoPh.223..643H}, and HLLD
\citep{2009MNRAS.393.1141M} can be adopted to compute numerical fluxes
for magnetofluids. 
For most of cases, we utilize HLL scheme, but we show 1-dimensional
numerical tests with HLL, HLLC and HLLD scheme in \S~\ref{mhd}.

For the 3rd step, we solve equation (\ref{eq:dadvection}) 
using numerical fluxes $f_{i\pm 1/2}$ and obtain auxiliary conserved
variables, 
$\mathcal{U}^* =(D^*, {\bmath m}^*, e^*, {\bmath B}^*, E^*, {\bmath F}_r^*)$,
where superscript of asterisk indicates that 
the quantity is computed at the 3rd step.
By the procedures so far, the advection terms are already solved,
and the gas-radiation interaction (equation [\ref{eq:dsource}]) 
remains only.
Hence, we obtain two of the conserved variables at the next timestep,
$D^{n+1}=D^*$ and ${\bmath B}^{n+1}={\bmath B}^{*}$.
In addition, although gas-radiation interaction changes energy 
and momentum for the gas and radiation, total energy and momentum 
of radiation magnetofluids in a local grid are conserved.
It implies that the total energy and the total momentum at the next timestep,
$e_t^{n+1}$ and ${\bmath m}_t^{n+1}$,
are obtained as,
\begin{eqnarray}
 e_\mathrm{t}^{n+1} &=&e^{*} + E_r^{*},\label{eq:Etot}\\
  \bmath m_\mathrm{t}^{n+1}&=&\bmath m^{*} + \bmath F_r^{*}\label{eq:mtot}.
\end{eqnarray}

For the 4th step, we calculate
${\bmath m}^{n+1}$, $e^{n+1}$, $E_r^{n+1}$, and ${\bmath F}_r^{n+1}$.
In particular, we calculate $E_r^{n+1}$ and ${\bmath F}_r^{n+1}$
by solving the gas-radiation interaction [equation (\ref{eq:dsource})],
and ${\bmath m}^{n+1}$ and $e^{n+1}$ are evaluated by
${\bmath m}^{n+1}={\bmath m}_t^{n+1}-{\bmath F}_r^{n+1}$
and $e^{n+1}=e_t^{n+1}-E_r^{n+1}$.
The primitive variables at the next timestep,
$\mathcal{P}^{n+1}$, is simultaneously computed.

In this step, we iteratively solve the equation (\ref{eq:dsource})
for radiation energy density and the radiation flux.
The source terms include 
primitive variables of fluids,
$\mathcal{S}=\mathcal{S} (E_r, F^i_r, D^{ij}, \mathcal{P}_h, {\bmath
B})$,
where $\mathcal{P}_h^{(m)}$ represents primitive variables of fluids
(i.e., $\rho, \bmath u$ and $p_g$).
We evaluate $E_r$ and $F_r^i$ at $(m+1)$-step in an implicit manner
with using $D_r^{ij}$ and $\mathcal{P}_h$ at $(m)$-step as
\begin{equation}
 \mathcal{U}^{(m+1)} = \mathcal{U}^{*} + \Delta t \mathcal{S}(E_r^{(m+1)},
  F_r^{j,(m+1)}, D_r^{jk,(m)}, \mathcal{P}_h^{(m)}, \bmath B^{n+1}).
  \label{eq:simp}
\end{equation}
The explicit form of this equation is represented later.
After solving the equation (\ref{eq:simp}), 
we calculate 
\begin{eqnarray}
 e^{(m+1)} &=& e_\mathrm{t}^{n+1} - E_r^{(m+1)},\label{eq:efromEtot}\\
 \bmath m^{(m+1)} &=& \bmath m_\mathrm{t}^{n+1} - \bmath F_r^{(m+1)}. 
  \label{eq:mfrommtot}
\end{eqnarray}
Since all the conserved variables at $(m+1)$-step are obtained,
we recover primitive variables $\mathcal{P}^{(m+1)}$ 
from $\mathcal{U}^{(m+1)}$ 
(the recovery method is mentioned later).
Then we again solve equation (\ref{eq:simp})
using updated primitive variables 
$\mathcal{P}_h^{(m+1)}$ and 
$D^{jk,(m)}=D^{jk}(E_r^{(m+1)},\bmath F_r^{(m+1)})$ 
\citep[a similar method is found in relativistic resistive MHD by][]{2009MNRAS.394.1727P}.
By setting $\mathcal{P}_h^{(0)}$ to be $\mathcal{P}_h^n$
and $D_r^{ij,(0)}$ to be $D_r^{ij}(E_r^{(n)},\bmath F_r^{(n)})$,
we continue the iteration until successive variables $\Delta
E_r^{(m+1)}\equiv E_r^{(m+1)}-E_r^{(m)}$,
$\Delta F_r^{(m+1),i}\equiv F_r^{(m+1),i}-F_r^{(m),i}$, and
$\delta P_h^{(m+1)}\equiv P_h^{(m+1)} - P_h^{(m)}$ fall below a
specified tolerance. When solutions converge, we apply them to the
solutions at $n+1$ timestep ($\mathcal{P}^{n+1}=\mathcal{P}^{(m+1)}$).

An explicit form of equation (\ref{eq:simp}) for the radiation field
is given by 
\begin{eqnarray}
 \bmath C^{(m)}
 \left(\begin{array}{c}
  E_r^{(m+1)}\\
  F_r^{k,(m+1)}\\
  \end{array}\right)=
 \left[\begin{array}{c}
   E^*_r + \left(4\pi\rho\gamma\kappa \mathrm{B}\right)^{(m)}\Delta t\\
   	F^{j,*}_r +\left(4\pi\rho u^{j}\kappa \mathrm{B}\right)^{(m)}\Delta t
        \end{array}\right].\label{eq:imp2_matlin}
\end{eqnarray}
where
\begin{eqnarray}
\bmath C \equiv \bmath 1 - \Delta t \bmath X,
\end{eqnarray} 
and
\begin{eqnarray}
\bmath X &=& \left(\begin{array}{cc}
	    X_{11}, X_{12}\\
            X_{21}, X_{22}
		 \end{array}
			\right),\nonumber \\
X_{11} &=& \rho  \gamma
 \left[-\kappa 
  + \sigma_s \left(|\bmath u|^2 +u_p u_q D_r^{pq}\right)
 \right]
 \nonumber \\
X_{12} &=& \rho u_k
 \left[\kappa-\sigma_s\left(\gamma^2 + |\bmath u|^2\right)\right]
 \nonumber \\
X_{21} &=& \rho 
 \left[k u_p D_r^{jp} 
  + \sigma_s u^j\left(\gamma^2 + u_p u_q D_r^{pq}\right)\right]
\nonumber \\
X_{22} &=& -\rho\gamma
 \left[(\kappa+\sigma_s) \delta^{j}_k +2 u^j u_k\right]
\end{eqnarray}
By inverting $4\times 4$ matrix $\bmath C$ directly, we obtain conserved
variables $E_r^{(m+1)}$ and $F_r^{j,(m+1)}$.

Here we mention the recovery method for 
converting from the conserved variables
to the primitive variables.
By the 3rd step, $D^{n+1}$ and ${\bmath B}^{n+1}$ are obtained 
as we have already mentioned.
In the 4th step, we have ${\bmath m}^{(m+1)}$ and $e^{(m+1)}$
by solving equations (\ref{eq:simp}-\ref{eq:mfrommtot}).
Then, three unknown variables $\rho^{(m+1)}, \bmath u^{(m+1)},
p^{(m+1)}_g$ are computed 
by solving 
a single non-linear equation $g(W)=0$ on $W=\rho h \gamma^2$
using Newton-Raphson method,
\begin{eqnarray}
 g(W) &=& W - p_g + \left(1-\frac{1}{2\gamma^2}\right)|\bmath B|^2
  - \frac{S^2}{2W}-e,\label{eq:toprimr2mhds}\\
 \gamma &=& \left[1-\frac{S^2(2W+|\bmath B|^2)+|\bmath m|^2 W^2}
	  {(W+|\bmath B|^2)^2 W^2}\right],\\
 p_g &=& \frac{W-D\gamma}{\Gamma_1 \gamma^2},\\
 \frac{d\gamma}{dW}&=&-\frac{\gamma^3}{W^3\left(W+|\bmath B|^2\right)}
  \nonumber \\
 &\times& \left[
   |\bmath m|^2 W^3 + 3 S^2W(W+ |\bmath B|^2) + S^2 |\bmath B|^4
\right],\\
 \frac{d p_g}{d W} &=& \frac{\gamma\left(1+D\frac{d\gamma}{dW}\right)
  -2W\frac{d\gamma}{dW}}{\Gamma_1 \gamma^3},\\
 \frac{dg}{dW}&=&1-\frac{dp_g}{dW}+\frac{|\bmath
  B|^2}{\gamma^3}\frac{d\gamma}{dW}
  +\frac{S^2}{W^3},\label{eq:toprimr2mhde}
\end{eqnarray}
where $\Gamma_1 = \Gamma/(\Gamma-1)$, and $S=\bmath m \cdot \bmath
B$ \citep{2006MNRAS.368.1040M, 2007MNRAS.378.1118M}.
This recovery method in SR-RMHD is the same
with that in relativistic pure MHD.

We noted that our scheme does not guarantee the physical 
constraint $|\bmath F_r| \leq E_r$. If a truncation error leads to 
$|\bmath F_r| > E_r$, unphysical solutions appear. 
To avoid this problem, we artificially
reduce the radiation flux without changing the direction of radiation
flux if the condition is violated, as 
\begin{equation}
 \bmath F_r \rightarrow \bmath F_r \mathrm{min}
\left(1, \frac{E_r}{|\bmath F_r|}\right).
 \label{eq:Frmodify}
\end{equation} 
 We confirmed that $|\bmath F_r|$ rarely exceeds $E_r$ and above 
procedure is applied in the test problems described in section 4.

\subsection{SR-R2MHD}\label{numR3MHD}
In SR-R2MHD, we solve equations (\ref{geq:ampere}),
(\ref{geq:resistive})-(\ref{geq:charge}), and
(\ref{eq:r2mcons})-(\ref{eq:r2fr})
so that we have 16 hyperbolic
equations. Note that equation (\ref{geq:ampere}) becomes stiff for the
ideal limit ($\eta \rightarrow 0$). Thus we solve SR-R2MHD equations 
using operator splitting as well as in SR-RMHD.
The primitive variables ($\mathcal{P}$), conserved variables
($\mathcal{U}$), fluxes ($\mathcal{F}$) and source terms ($\mathcal{S}$)
for SR-R2MHD are given by
\begin{equation}
 \mathcal{P} = 
  \left(\begin{array}{c}
   \rho \\
   u^j \\
   p_g \\
   B^j \\
   E^j\\
   \rho_e\\
   E_r \\
   F_r^j\\
\end{array}\right),\ 
 \mathcal{U} 
  = \left(\begin{array}{c}
     D \\
     m^j\\
     e\\
     B^j\\
     E^j\\
     \rho_e\\
     E_r\\
     F_r^j\\
  \end{array}\right),\ 
  \mathcal{F}^k
  = \left(\begin{array}{c}  
     D v^k\\
     \Pi^{jk}\\
     m^k \\
     \varepsilon^{jkl} E_l\\
     -\varepsilon^{jkl} B_l\\
     j^k\\
     F_r^j \\
     P_r^{jk}
   \end{array}\right),
\end{equation}
and 
\begin{eqnarray}
 \mathcal{S} &\equiv&\mathcal{S}_a + \mathcal{S}_b\nonumber \\
 &=&
  \left(\begin{array}{c}
     0 \\
     0\\
     0^j\\
     0\\
     -qv^j\\
      0\\
     0\\
     0\\
	\end{array}\right) +
  \left(\begin{array}{c}
     0 \\
     -S_E\\
     E_F^j\\
     0\\
     -\frac{\gamma}{\eta}\left[E^j + \varepsilon^j_{\ lm}v^l B^m - v^j v_l E^l\right]\\
     0\\
     S_E\\
     S_F^j\\
  \end{array}\right) 
\label{eq:r3source}
\end{eqnarray}
Here we decompose $\mathcal{S}$ as $\mathcal{S}=\mathcal{S}_a +
\mathcal{S}_b$. Note that $\mathcal{S}_b$ makes equation
(\ref{eq:dsource}) stiff
for the ideal limit ($\eta\rightarrow 0$) or when the cooling/scattering
time scale is shorter than the dynamical time scale. 
On the other hand, $\mathcal{S}_a$ is independent of $\eta$, $\kappa$
and $\sigma_s$ so
that we can integrate this term explicitly \citep{2007MNRAS.382..995K,
2009MNRAS.394.1727P}. Then, one-dimensional discretization of equation
(\ref{eq:1dform}) is given by
\begin{eqnarray}
 \mathcal{U}^*_i = \mathcal{U}^n_i -
  \frac{\Delta t}{\Delta x}\left(f_{i+1/2}^{n} -
			    f_{i-1/2}^n\right) + \mathcal{S}_{a,i} \Delta t,
  \label{eq:dadvection2}\\
 \mathcal{U}^{n+1}_i = \mathcal{U}^*_i
  +\mathcal{S}_{b,i}^{n+1}\Delta t,\label{eq:dsource2}
\end{eqnarray}

For the 1st step, we compute surface values of primitive variables. The
procedure is the same with that for SR-RMHD given in equations
(\ref{eq:phalf})-(\ref{eq:dp}).

For the 2nd step, we compute numerical fluxes $f_{i\pm 1/2}$ using HLL
scheme. Similar to SR-RMHD, eigenvalues and eigenvectors can be computed
independently for the electromagnetofluids and radiation.
For a fluid, a fastest wave speed is a light speed since we
solve a full set of Maxwell equations. Thus we take $\lambda^{\pm}_m=\pm 1$
in equation (\ref{eq:hllf}) so that the HLL
scheme reduces to the Lax-Friedrich scheme \citep{2007MNRAS.382..995K} .
For the radiation field, we can compute numerical fluxes using HLL
scheme as described in the previous section.

For the 3rd step, we solve equation (\ref{eq:dadvection2}) using
numerical fluxes. At this step, $\mathcal{S}_a$ is integrated explicitly. 
The conserved variables at
the auxiliary step $\mathcal{U}^*$ is then obtained.
As discussed in Section \ref{numR2MHD}, 
we can compute total energy ($e_t = e + E_r$) and momentum ($m_t^j = m^j +
F_r^j$) at $n+1$-time step from equations
(\ref{eq:Etot})-(\ref{eq:mtot}) after the 3rd step, i.e., before solving
equation (\ref{eq:dsource2}).

For the 4th step, we integrate source terms $\mathcal{S}_b$, which
consist of two equations, the radiation moment equations and the
Ampere's law. Since
$E_r$ and $\bmath F_r$ do not appear in $\bmath j$, we can integrate equation
(\ref{eq:dsource2}) for radiation fields and electric fields
independently.

For radiation moment equations, we can integrate $\mathcal{S}_b$
using implicit scheme described in the previous subsection.
For the Ampere's law, we adopt an implicit scheme proposed by
\cite{2009MNRAS.394.1727P}. In their scheme, the electric field is
obtained by analytically inverting $3\times 3$ matrix:
\begin{eqnarray}
 E^{(m+1),j} &=& \frac{(\gamma^{(m)} +\Delta t/\eta)\delta^{jk}
  -u^{(m),j} u^{(m),k}\Delta t/\eta}
  {(1+\gamma^{(m)}\Delta t/\eta)(\gamma^{(m)} + \Delta t/\eta)}
  \nonumber \\
 &\times&\left[E^{*,k}-\varepsilon^k_{pq} u^{(m),p} B^{n+1,q}\Delta t/\eta\right],\label{eq:imp_E}
\end{eqnarray}
As in implicit integration for radiation moment equation, $\bmath v$ is
evaluated at $(m)$-th iteration step.

Then, we recover $\mathcal{P}^{(m+1)}$ from
$\mathcal{U}^{(m+1)}$. The gas energy
density $e_h$ and momentum $\bmath m_h$ for fluids are computed from 
equations (\ref{eq:Etot})-(\ref{eq:mtot}) by
\begin{eqnarray}
 e_h^{(m+1)} &=& \rho h \gamma^2 - p_g \nonumber \\
 &=&  
  e_t^{n+1} - E_r^{(m+1)} 
  - \frac{|\bmath B^{n+1}|^2 + |\bmath E^{(m+1)}|^2}{2},\\
 \bmath m_h^{(m+1)} &=& \rho h \gamma \bmath u \nonumber \\
&=&  \bmath m_t^{n+1} - \bmath F_r^{(m+1)} 
  - \bmath E^{(m+1)}\times \bmath B^{n+1}.
\end{eqnarray}
Note that we have $\bmath E$ at $(m+1)$-th iteration step since it is
both primitive and conserved variables in SR-R2MHD.
Thus the electromagnetic energy density and Poynting flux at $(m+1)$-step
are already determined.  
We compute $\rho, u_r, p_g$ from $D, \bmath m_h, e_h$
while they are computed from $D, \bmath m, e$ in SR-RMHD.
\begin{deluxetable*}{lccccccccccccc}
\tabletypesize{\scriptsize}
\tablecaption{List of Simulation Runs}
\tablewidth{0pt}
\tablehead{
\colhead{model} & \colhead{$\kappa$} &\colhead{$\Gamma$}
 &\colhead{state} & \colhead{$\rho$} &
 \colhead{$p_g$} & \colhead{$u^x$} & \colhead{$u^y$}
 & \colhead{$u^z$} & \colhead{$B^x$} &
\colhead{$B^y$} & \colhead{$B^z$} & \colhead{$E_r'$}
}
\startdata
 Contact Wave &0 &$\frac{5}{3}$& 
 L & 10     & 1   & 0 & 1.02 & 0.292 & 5   & 1  & 0.5 & 0\\
   \          &\ &\ & 
 R & 1      & 1   & 0 & 1.02 & 0.292 & 5   & 1  & 0.5 & 0\\
 Rotational Wave &0 &$\frac{5}{3}$& 
 L & 1     & 1   & 0.566 & -0.424 & 0.707 & 2.4   & 1  & -1.6 & 0\\
 \          &\ &\ & 
 R & 1      & 1   & 0.566 & -0.723 & 0.636 & 2.4   &-0.1  & -2.18 & 0\\
 MHDST1       &0 &2& 
 L & 1      & 1   & 0 & 0    & 0     & 0.5 & 1  & 0   & 0\\
   \          &\ &\ & 
 R & 0.125  & 0.1 & 0 & 0    & 0     & 0.5 & -1 & 0   & 0\\
 MHDST2       &0 &$\frac{5}{3}$ & 
 L & 1      & 0.1 & 22.3 & 0    & 0     & 10 & 7  & 7 & 0\\
   \          &\ &\ & 
 R & 1      & 0.1 & 22.3 & 0    & 0     & 10 &-7 &
 -7 & 0\\
 RHDST1       &0.4 &$\frac{5}{3}$ & 
 L & 1.0      & $3.0\times 10^{-5}$ & 0.015 & 0    & 0     & 0 & 0  & 0
 & $1.0\times 10^{-8}$\\
   \          &\ &\ & 
 R & 2.4      & $1.61\times 10^{-4}$ & $6.25\times 10^{-3}$ & 0    & 0
 & 0 &0 & 0 & $2.50\times 10^{-7}$\\
 RHDST2       &0.3 &2& 
 L & 1.0      & 60 & 10 & 0    & 0     & 0 & 0  & 0
 & 2\\
   \          &\ &\ & 
 R & 8      & $2.34\times 10^{3}$ & 1.25 & 0    & 0
 & 0 &0 & 0 & $1.13\times 10^{3}$\\
 RHDST3       &0.08 &$\frac{5}{3}$& 
 L & 1.0      & $6.0\times 10^{-3}$ & 0.69 & 0    & 0     & 0 & 0  & 0
 & 0.18\\
   \          &\ &\ & 
 R & 3.65      & $3.59\times 10^{-2}$ & 0.189 & 0    & 0
 & 0 &0 & 0 & $1.30$\\
\enddata
\tablecomments{Parameter sets of numerical tests. Scattering coefficient
 $\sigma_s$ is taken to be zero in all models.}
\label{tab:testparam}
\end{deluxetable*}

We adopt a recovery method developed by
\cite{2009ApJ...696.1385Z}. In their method, a single quartic equation
on $u=\sqrt{|\bmath u|^2}$ is numerically solved:
\begin{eqnarray}
&&\Gamma_1^2\left(e_h^2 - |\bmath m_h|^2 \right) u^4
- 2 \Gamma_1 |\bmath m_h| D u^3 \nonumber \\
&&+ \left[\Gamma_1^2 e_h^2 - D^2 - 2 \Gamma_1(\Gamma_1-1)|\bmath m_h|^2\right]
u^2\nonumber \\
&& -2(\Gamma_1-1)D |\bmath m_h| u - (\Gamma_1-1)^2 |\bmath m_h|^2=0,
  \label{eq:primquart}
\end{eqnarray}
where $\Gamma_1=\Gamma/(\Gamma-1)$. $p_g$, $\rho$ and $u^i$ are
obtained by
\begin{equation}
 p_g = \frac{1}{\Gamma_1-1}\left(\frac{|\bmath m_h|}{\gamma \sqrt{\gamma^2-1}}-\frac{D}{\gamma-1}\right),
\end{equation}
\begin{equation}
 \rho = \frac{D}{\gamma},
\end{equation}
and
\begin{equation}
 \bmath u = \frac{\bmath m_h}{\rho h \gamma^2}.
\end{equation}
Here we omit a superscript ${(m)}$ for simplicity. 

Now, we obtain all of primitive variables at $(m+1)$-th step.
Similar to SR-RMHD, $\mathcal{P}_h$ are evaluated at $(m)$-th step when
equation (\ref{eq:dsource2}) is solved.
Thus, we again solve equation (\ref{eq:dsource2}) using updated
$\mathcal{P}_h$. This iteration is continued
until successive variables fall below a specified tolerance. 

\section{Numerical Tests}\label{test}
In this section, we show results of some numerical tests for one- and
two-dimensional problems. 
Results for one-dimensional problems of relativistic pure
MHD are shown in \S~\ref{mhd}.
We present results of one- and two-dimensional problems of 
propagating radiation energy in \S~\ref{rad}
and one-dimensional shock tube problems of relativistic radiation
hydrodynamics in \S~\ref{rad-HD}.
In \S~\ref{r3recon},
we attempt the relativistic magnetic reconnection problem
by our SR-R2MHD code.

\begin{figure*}
 \begin{center}
  \includegraphics[width=12cm]{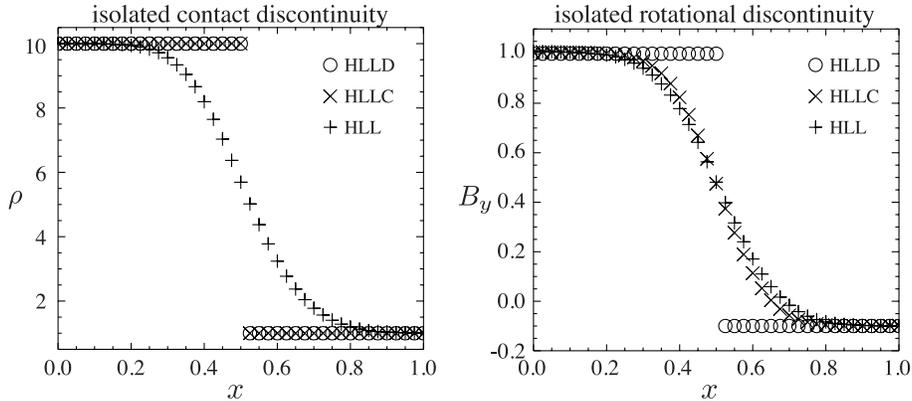}
  \caption{Results for the isolated stationary contact (left) and
  rotational (right) discontinuities. Density and $B_y$ are shown,
  respectively. Plus signs, crosses, and open circles correspond
  to results with HLL, HLLC, and HLLD scheme, respectively. The number of
  grid points is $N_x = 40$.}
  \label{fig:icrd}
 \end{center}
\end{figure*}
\begin{figure*}
 \begin{center}
  \includegraphics[width=18cm]{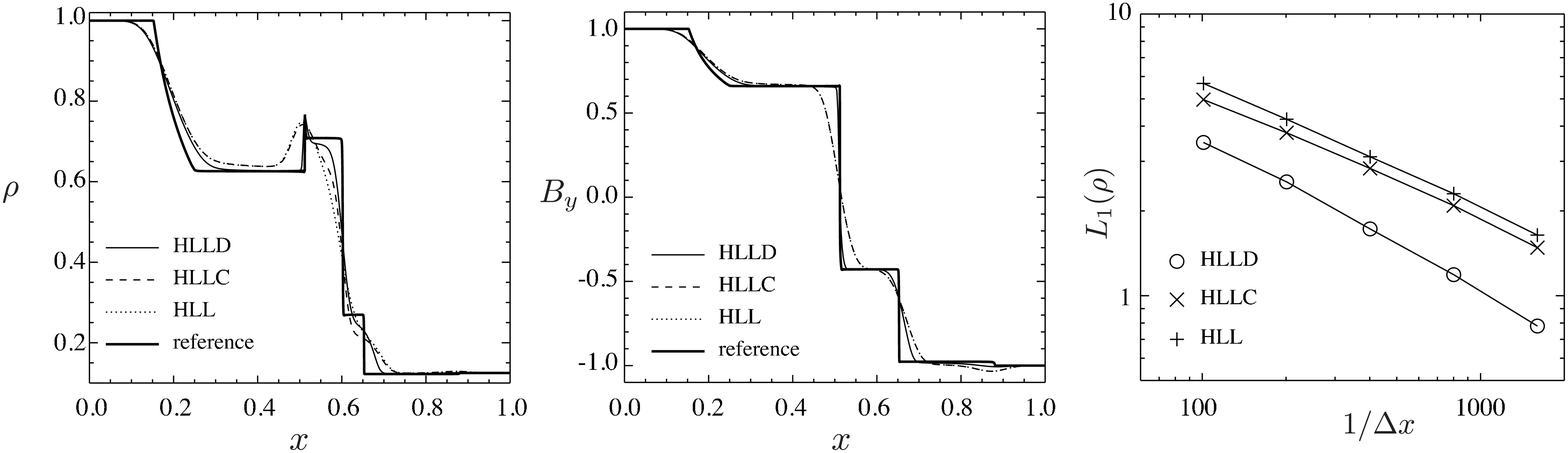}
  \caption{Results for model MHDST1 at $t=0.4$. Density
  and $B_y$ are plotted in the left and central panels,
  respectively. Solid, dashed and dotted curves denote results
  of the first order HLLD, HLLC, and HLL schemes with $N_x=400$, while
  thick solid curves do those for the second order HLLD scheme with
  $N_x=6400$ (reference solution), respectively. A right panel shows
  the $L_1(\rho)$ norm compared with reference solutions. Open
  circles, crosses, and plus signs correspond to those for HLLD,
  HLLC, and HLL schemes, respectively.}
  \label{fig:mhdshtb1}
 \end{center}
\end{figure*}
\begin{figure*}
 \begin{center}
  \includegraphics[width=18cm]{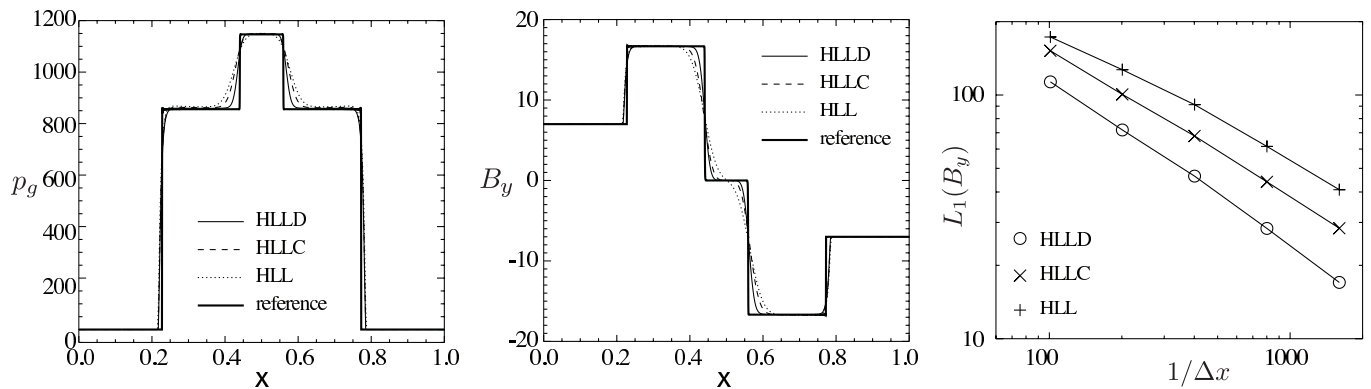}
  \caption{Results for model MHDST2 at $t=0.4$. The
  gas pressure and $B_y$ profiles are plotted in the left and central
  panels, respectively. Solid, dashed and dotted curves denote 
  results of the first order HLLD, HLLC, and HLL schemes with $N_x=400$,
  while thick solid curves do those for the second order HLLD scheme with
  $N_x=6400$ (reference solution), respectively. A right panel shows
  the $L_1(B_y)$ norm compared with reference solutions. Open
  circles, crosses, and plus signs correspond to those for HLLD,
  HLLC, and HLL schemes, respectively.}
  \label{fig:mhdshtb2}
 \end{center}
\end{figure*}
\subsection{Relativistic Ideal Magnetohydrodynamics}\label{mhd}
We perform four numerical tests of one-dimensional shock tube problems
without radiation and resistivity.
An initial discontinuity is situated at $x=0.5$
in a computational domain of $x=[0, 1]$.
Initial states of left ($x<0.5$) and right ($x>0.5$) regions
for each problem are listed in Table~\ref{tab:testparam}.

In the following subsection, relativistic MHD equations are solved using
a 1st order accurate scheme in space. 
Numerical fluxes are computed by HLL
\citep{1983siamRev...25..35..61}, HLLC \citep{2006MNRAS.368.1040M}, and
HLLD \citep{2009MNRAS.393.1141M} scheme. 
We note that the results by the alternative HLLC scheme
\citep{2007JCoPh.223..643H}
are consistent with that by the scheme of \cite{2006MNRAS.368.1040M}.

An accuracy of our numerical code is verified by calculating 
$L-1$ norm:
\begin{equation}
 L_1(g) = \sum_{i=1}^{N_x} |g^{ref}_i - g_i| \Delta x_i, \label{test:L1}
\end{equation}
where $N_x$ and $\Delta x_i$ are a number of grid points 
and a grid spacing. $g_i$ is a numerical solution of
some physical quantities, while $g^{ref}_i$ is a reference solution. 
We use numerical results of a second order HLLD scheme
with $N_x=6400$ for reference solutions, which are consistent with that
with more grids.
In this subsection, the CFL number is fixed to be 0.8.

\subsubsection{Isolated contact and rotational
   discontinuities}\label{ICRD}
For tests of stationary isolated contact and rotational discontinuities,
which are proposed by \cite{2009MNRAS.393.1141M},
we employ $N_x = 40$ and $\Gamma=5/3$. 
At the initial state, there is a density jump, while the other
quantities are continuous for the isolated contact wave problem
(see Contact Wave in Table~\ref{tab:testparam}). 
The velocity and  magnetic field vectors are discontinuous, while $\rho$ and
$p_g$ are invariant for the rotational discontinuity
(see Rotational Wave in Table~\ref{tab:testparam}). 

In Figure~\ref{fig:icrd},
we plot the density, $\rho$, 
for isolated contact discontinuity (left panel)
and $y$-component of magnetic fields, $B_y$,
for isolated rotational discontinuity (right panel)
at $t=1.0$. 
Plus signs, crosses, and open circles denote results with HLL,
HLLC, and HLLD solver, respectively. 

We find in the left panel that HLLC and HLLD schemes,
which intrinsically capture an entropy wave, 
can reproduce the contact surface, while a density
profile becomes smoothed out 
in the case of HLL scheme.
The right panel clearly shows that HLLD scheme, 
which can intrinsically capture the Alfv\'en wave, 
recovers a surface of the rotational discontinuity, 
in contrast to HLLC and HLL schemes.
Here, we note that the profile of $B_y$
is slightly steeper by HLLC scheme 
than by HLL scheme at $x\sim 0.5$.
This is because the numerical viscosity 
is smaller in HLLC scheme than in the HLL scheme.
We recognize that our numerical code can capture the entropy wave 
by the HLLC and HLLD schemes, and Alfv\'en waves by HLLD scheme correctly.

\subsubsection{MHD Shock Tube 1}\label{mhdshtb1}
The relativistic extension of the shock tube problem
by \citet{1988JCoPh..75..400B}
is proposed by many authors \citep{2001ApJS..132...83B,
2003A&A...400..397D, 2006MNRAS.368.1040M, 2009MNRAS.393.1141M}. 
In this problem (model MHDST1), an initial
discontinuity is broken up into a fast rarefaction wave, a compound wave,
a contact discontinuity, a slow shock and a fast rarefaction wave, 
from left to right.

Figure~\ref{fig:mhdshtb1} shows numerical results at $t=0.4$. 
Here, we employ $\Gamma=2$.
Left and central panels show $\rho$ and $B_y$ profiles with $N_x=400$. 
Solid, dashed, and dotted curves denote results
of 1st order HLLD, HLLC, and HLL schemes, while
reference solutions are plotted by thick solid curves.
Although
the profile of the rarefaction wave front
($x\sim 0.1-0.3$)
is almost independent of solvers,
HLLD scheme only gives improved profiles
at the compound wave ($x\simeq 0.5$), the contact
surface ($x \simeq 0.6$), and the slow shock ($x\simeq 0.65$)
\citep[see also Fig. 3 in][]{2009MNRAS.393.1141M}.

The right panel of Fig.\ \ref{fig:mhdshtb1} shows a $L_1$ norm of $\rho$
at $t=0.4$. We can see that the error linearly decreases with
decreasing a grid size in all schemes. 
We note that HLLD scheme drastically reduces 
numerical errors compared to other solvers. 
We find $L_1^\mathrm{HLL} : L_1^{HLLC}: L_1^{HLLD}=
1.0 : 0.91 : 0.55$,
where superscripts of $L_{\rm 1}$ indicate the scheme.
Here, note that the HLLD scheme takes a longer computational time. 
We find 
$t_\mathrm{HLL} : t_\mathrm{HLLC} : t_\mathrm{HLLD} = 1 : 1.27 : 1.61$, 
where $t_\mathrm{HLL}$, $t_\mathrm{HLLC}$, and $t_\mathrm{HLLD}$
are computational time by HLL, HLLC, and HLLD scheme.
\begin{figure*}
 \begin{center}
  \includegraphics[width=18cm]{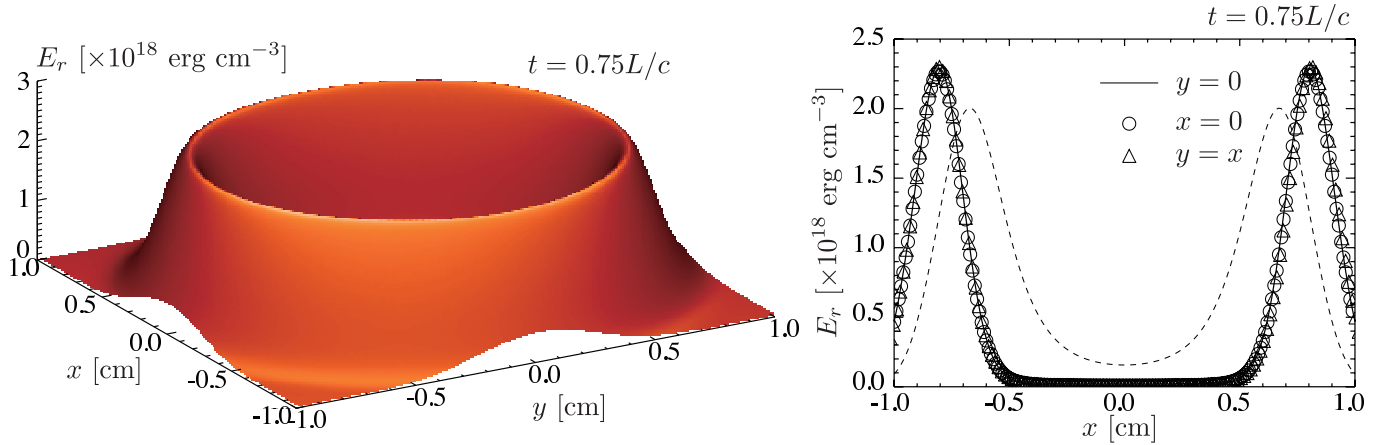}
  \caption{(left) bird's eye view of $E_r$ at $t=0.75L/c$. 
(right) 1-dimensional profiles of $E_r$ at $t=0.75L/c$. Solid curve
  denotes profiles on $y=0$, while open circles and open triangles
  denote $E_r$ on $x=0$ and $y=x$, respectively. 
  Dashed curve also shows $E_r(x,y=0)$ but we initially set $F_r^i=0.$
}
  \label{fig:caldera2d}
 \end{center}
\end{figure*}
\subsubsection{MHD Shock Tube 2}\label{mhdshtb2}
We perform a test calculation for a collision of oppositely directing
relativistic flows \citep[model
MHDST2,][]{2001ApJS..132...83B, 2003A&A...400..397D,
2006MNRAS.368.1040M, 2009MNRAS.393.1141M}. 
Here, the bulk Lorentz factor is $\gamma\simeq 22.4$,
and we set the specific heat ratio, $\Gamma$, to be $5/3$.

Figure ~\ref{fig:mhdshtb2} shows results at $t=0.4$. 
Left and central panels indicate profiles of $p_g$ and $B_y$. 
Solid, dashed, and dotted curves represent results
of first order HLLD, HLLC, and HLL schemes with $N_x=400$, while
a reference solution (a second order HLLD scheme with $N_x=6400$)
is shown by thick solid curves.

We find in Figure ~\ref{fig:mhdshtb2} that 
two slow mode shocks ($x\sim 0.45$, $0.55$)
are sandwiched by two fast mode shocks ($x\sim 0.2$, $0.8$) 
and that all of approximate Riemann solvers we adopted can
capture the fastest mode (fast magnetosonic wave). 
Note that although a slow mode is not taken into account in HLLD scheme,
less numerical viscosity leads to optimal solution.

The $L_1$ norm for $B_y$ is shown in the right panel of
Figure~\ref{fig:mhdshtb2}. Filled circle, crosses, and plus signs indicate
results with HLLD, HLLC, and HLL solver, respectively. 
The error linearly decreases with the grid spacing for all numerical
schemes and the HLLD scheme is approved as the best numerical scheme
in comparison with the other approximate Riemann solver. 
This panel also shows that the accuracy of HLLC scheme is better than
that of HLL scheme \citep[see also, Fig. 7 in][]{2009MNRAS.393.1141M}.
We find $L_1^\mathrm{HLL} : L_1^{HLLC}: L_1^{HLLD}=
1.0 : 0.74 : 0.51$,
and 
$t_\mathrm{HLL} : t_\mathrm{HLLC} : t_\mathrm{HLLD} = 1 : 1.11 : 1.57$.

In addition to the problems mentioned above
(Contact wave, Rotational wave, MHDST1, MHDST2), 
we performed several conventional
one dimensional relativistic shock tube problems proposed by authors, 
and demonstrate the relativistic self-similar expansion of
magnetic loop in two dimensions \citep{2011MNRAS.414.2069T}. 
We confirmed that our numerical code can pass these problems. 

\subsection{Tests for Propagating Radiation Energy}\label{rad}
We show results of numerical tests 
for propagating radiation energy. 
We recover a light speed $c$ in this subsection. 
Although we solve a full set of SR-RMHD equations throughout this subsection, 
the radiation hydrodynamics are not proofed.
The propagation of radiation energy in the static fluid 
is virtually tested,
since the radiation force as well as the gas pressure force 
is negligible and the velocity of the matter is almost kept null.

\subsubsection{Point Explosion}\label{explosion}
We show an expansion of radiation field from a point source.
We performed two-dimensional simulations in the $x-y$ plane with a
volume bounded by $x=[-L,L]$ and $y=[-L, L]$, where $L=1~\mathrm{cm}$. 
We use uniformly spaced grids of $200 \times 200$.
We assume a static ($v=0$)
and constant density profile with $\rho =
1~\mathrm{g~cm^{-3}}$. 
The absorption coefficient is given by
$\kappa=0.1~\mathrm{cm^2~g^{-1}}$,
and the scattering coefficient is set to be null.
Thus, a computational domain is optically thin, $\tau = \rho \kappa L=0.1$. 
The radiation energy and flux are initially given by
\begin{eqnarray}
 E_r = 10^{10} E_\mathrm{LTE} \exp\left[-\frac{x^2+y^2}{0.01}\right],\\
 F_x = cE_r\frac{x}{\sqrt{x^2+y^2}}\\
 F_y = cE_r\frac{y}{\sqrt{x^2+y^2}},
\end{eqnarray}
where $E_\mathrm{LTE}=a_RT_r^4=10^{10}~\mathrm{erg~cm^{-3}}$ is the 
radiation energy density at the local thermodynamic equilibrium 
(LTE, $T_g = T_r$). Here $T_r$ is the radiation temperature.
We solve the SR-RMHD equations with an first order
accuracy in space and time.

A left panel of Figure \ref{fig:caldera2d} 
shows bird's eye view of radiation energy density, $E_r$, at
$t=0.75L/c$. 
Also, one-dimensional profiles of $E_r$ on $x$-axis (solid curve),
$y$-axis (open circles), and $y=x$ (open triangles) are plotted in a
right panel. Following initial
enhancements of radiation energy, the radiation energy propagates 
in a circle with a light speed.
Since most of the radiation energy is transported 
without absorption by matter,
the radiation energy density 
decreases with a distance from the center, $r\equiv (x^2+y^2)^{1/2}$,
approximately as $E_r \propto r^{-1}$. 

Such a caldera structure also appears
even if we employ the Eddington approximation \citep{2013ApJ...764..122T}.
However, in this method, the wave front propagates 
with a speed of $c/\sqrt{3}$.
Such a reduction of the speed is induced by that 
the radiation field is assumed to be isotropic in Eddington
approximation ($D_r^{ij}=1/3$).
On the other hand, 
the Eddington tensor is given 
by taking account of the non-isotropic radiation fields
in the M-1 closure
(see Equations [\ref{geq:DtensM1}]-[\ref{geq:DtensM1end}]).
\citet{2001ApJS..135...95T} attempted a similar test problem
with using FLD approximation.
They also succeeded in reproducing the propagating radiation energy 
with speed of light.
However, since the radiation flux is basically given 
by the gradient of the radiation energy density in FLD,
a caldera structure is not formed and 
the top-hat shaped distribution 
of the radiation energy density appears.
The M-1 closure has an advantage when 
the radiation transport in an optically thin medium is considered.

The right panel of Figure~\ref{fig:caldera2d} shows 1-dimensional
profiles of $E_r$ at $t=0.75L/c$. A solid curve denotes profiles at
$y=0$, while open circles and open triangles denote $E_r$ at $x=0$
and $y=x$, respectively. As discussed above, the radiation energy is
transported with the light speed and a caldera structure is formed. 
We can see that these three profiles are consistent, indicating that 
the space symmetry is assured in our numerical code. 

In this test problem, non-zero radiation fluxes are initially
given by $|\bmath F_r| = c E_r$. Then the radiation energy propagates
radially with light speed. 
When we take $F^i_r=0$ initially, the radiation energy
slowly expands compared to former results. This can be confirmed in the
right panel of Figure~\ref{fig:caldera2d}. A dashed curve shows
$E_r(x,y=0)$ at $t=0.75L/c$ for a model that $F_r^i$ is initially zero. 
We find that the wave front is slightly delayed for $|\bmath
F_r(t=0)|=0$ in comparison with that for $|\bmath F_r(t=0)|=c E_r$. 
This is because that the Eddington tensor becomes $1/3$ when $|\bmath F_
r|=0$, leading to the wave velocity of $c/\sqrt{3}$. Since the expansion
speed approaches to the light speed as time goes on, the gap of the wave
fronts does not widen furthermore. 
Note that the characteristic wave velocity remains $c/\sqrt{3}$ around
the origin via the nearly isotropic radiation fields. It makes profiles
more diffusive. Hence, the radiation energy density for the case of
$|\bmath F_r(t=0)|$=0 is not null around the origin.

\subsubsection{Beam}\label{beam}
We show a radiation transport with a certain angle to 
a grid \citep{2001A&A...380..776R, 2007A&A...464..429G}.
We used $400 \times 400$ grid points which cover the
computational domain $x=[0,L]$ and $y=[0,L]$ where $L=1~\mathrm{cm}$. 
We assume a constant density profile of $\rho = 1~\mathrm{g~cm^{-3}}$. 
We assume $\kappa = 1~\mathrm{g\ cm^{-3}}$ and $\sigma_s=0$, leading the
optical depth of $\tau = \rho \kappa L=1$. The radiation field is in LTE
with a matter, whose energy is $E_r = E_\mathrm{LTE}=
10^{10}~\mathrm{erg~cm^{-3}}$. 
We inject radiation from the boundary, $x=0$ and $y=[0.1,\ 0.2]$.
The injected radiation energy is $E_\mathrm{inj}=10^{5} E_\mathrm{LTE}$
and the radiation flux is given by $F_x = F_y = cE_r/\sqrt{2}$.
We adopted a free-boundary condition at the other boundaries. 
The SR-RMHD equations are solved with 2nd order accuracy in space and
time. 

\begin{figure}
 \begin{center}
  \includegraphics[width=8cm]{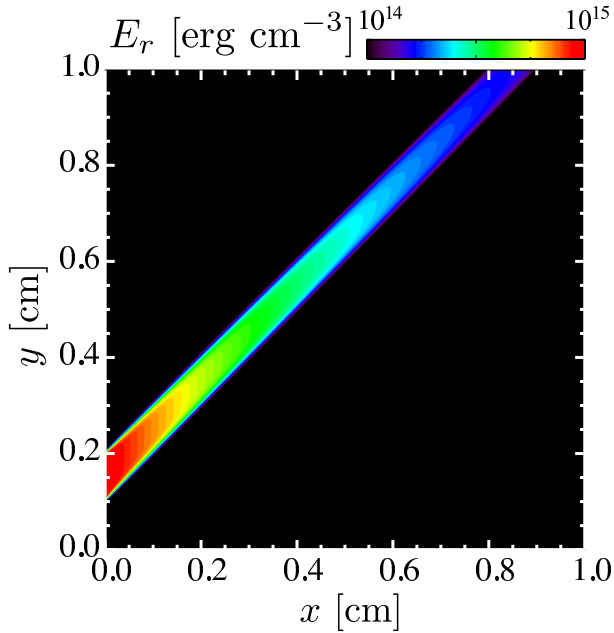}
  \caption{Snapshot of $E_r$ at $t=2 L/c$ for the beam test.}
  \label{fig:beam2d}
 \end{center}
\end{figure}

Figure ~\ref{fig:beam2d} shows a snapshot of $E_r$ at $t=2.0L/c$. 
We can see that a beam profile can be sharply captured in our
numerical scheme thanks to a 2nd order accurate scheme.
If we employ the Eddington approximation, 
the beam would broaden since the isotropic radiation field is assumed
(this point will be discussed again in \S\ref{shadow}).
Also, the beam can not be reproduced in principle
in the case of FLD approximation.
Since the radiation flux is given as a function of the 
the gradient of $E_r$,
the radiation energy propagates in a circle.

\begin{figure}
 \begin{center}
  \includegraphics[width=8cm]{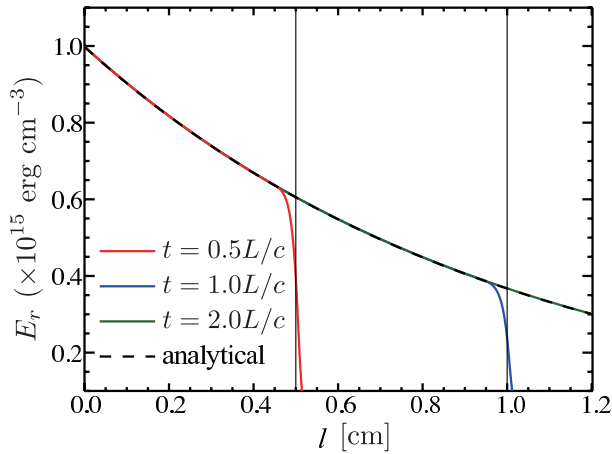}
  \caption{Profiles of the radiation energy along the beam
  at $t=0.5L/c$ (red), $1.0L/c$ (blue) and $2.0L/c$ (green). 
  Thin solid curves denote the wave front at $t=0.5L/c$ and $t=L/c$,
  while a black dashed curve represents an analytical solution of steady
  model.}
  \label{fig:beam1d}
 \end{center}
\end{figure}

We plot in Figure \ref{fig:beam1d} 
that the radiation energy density along the
beam ($y=x+0.15$) at $t=0.5, 1.0, 2.0 L/c$.
Here $l$ is a distance from 
the center of the injection point [$(x,\ y)=(0,\ 0.15)$]. 
The radiation front, which propagates with a speed of light, 
can be excellently captured thanks to the 2nd order accurate scheme.

Since the radiation energy is absorbed by the matter,
and since the emission of the matter is negligible,
the radiation energy density decreases with an increase of $l$.
Then, the profile of $E_r$ is analytically expressed as
\begin{equation}
 E_r(l) = E_\mathrm{inj} \exp(-\rho \kappa l), \label{eq:dampM1}
\end{equation}
within the wave front of $l=ct$ (a dashed curve).
We can see that our numerical results excellently recover 
the analytical solution. 

\subsubsection{Shadow}\label{shadow}
We show the light propagation around a dense matter.
This test was proposed by \cite{2003ApJS..147..197H} and
\cite{2007A&A...464..429G} adopted the M-1 formulation to the problem.
We perform simulations with M-1 closure and with Eddington approximation.

We utilize a simulation box bounded by $x=[-5, 10]~\mathrm{km}$ and
$y=[0, 5]~\mathrm{km}$ with grid
points of $300\times 100$. 
By setting $C_\mathrm{cfl}=0.5$,
the timestep is 
$\Delta t=C_\mathrm{CFL} \Delta x/c=8.3\times 10^{-8} \mathrm{s}$.
We set $\sigma_s=0$ in the whole range of the domain.
We consider a dense clump embedded in the less dense matter.
The clump is located at the origin. 
The radius and the mass density are 
$r_0=2~\mathrm{km}$
and $\rho_1=30~\mathrm{g~cm^{-3}}$. 
The density of the surrounding rarefied matter
is set to be $\rho_0=10^{-6}~\mathrm{g~cm^{-3}}$. 
Since we here suppose $\kappa=10^{-2}~\mathrm{cm^2}~g^{-1}$,
the optical thickness of the clump
is $\rho_1\kappa r_0 = 6\times 10^{4}$,
and the less dense region is optically thin,
$\rho_1 \kappa \times 15 {\rm km} = 0.015$.
The radiation energy density is set to be constant,
$E_r=E_\mathrm{LTE}=10^{5}~\mathrm{erg~cm^{-3}}$, 
and the rarefied matter is LTE, initially.
We assume a uniform gas temperature with $T_g = T_r$ at the initial
state.


\begin{figure*}
 \begin{center}
 \includegraphics[width=18cm]{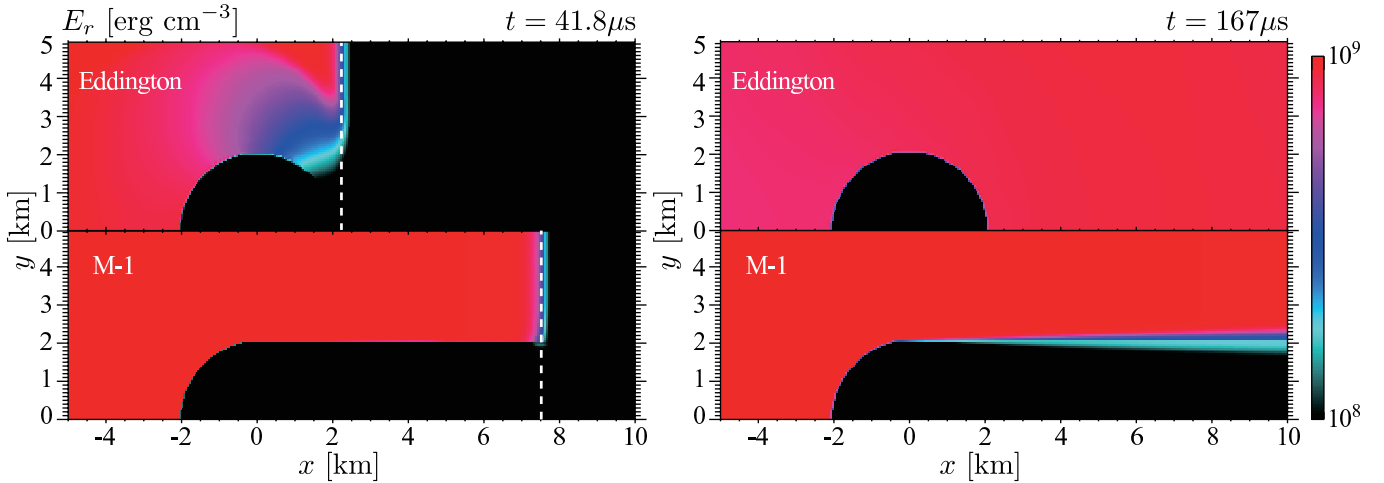}
  \caption{Color contours of $E_r$ at $t=41.8 \mu s$ and $167 \mu s$ for
 a shadow  test. White dashed lines in the left panel show wave fronts at
 $x=-0.5\ \mathrm{cm}+ct/\sqrt{3}$ with the Eddington approximation and
 $x=-0.5\ \mathrm{cm}+ct$ with the M-1 closure.}
  \label{fig:shadow1}
 \end{center}
\end{figure*}
The radiation is injected at the left boundary, $x=-5~\mathrm{km}$,
where the radiation energy density, $E_\mathrm{inj}$, is set to be 
$10^{4} E_\mathrm{LTE}$,
and the radiation flux is assumed as $F^x_{r}= cE_r$
with M-1 closure and $F^x_r=cE_r/\sqrt{3}$ with Eddington approximation.
A free boundary condition is employed at the upper ($x=10~\mathrm{km}$) 
and right ($y=5~\mathrm{km}$) boundaries.
At $y=0$, we use a symmetric boundary.

The radiation energy density 
at $t=41.8~\mathrm{\mu s}$ (left) and $167~\mathrm{\mu s}$ (right),
which correspond to $0.83$ and $3.3$ light
crossing time along the $x$-direction,
is presented by color contours in Figure \ref{fig:shadow1}.
Upper and lower panels are results with Eddington and M-1 
methods, respectively. 
In this Figure, 
we can see a shadow behind the clump in M-1 method ($x>2.0$ km). 
Since the parallel light injected from the left boundary
is absorbed by the dense clump,
and since the photons are not scattered ($\sigma_s=0$),
the lower right region of $x>2.0$ km and $y<2.0$ km
is darkened by shadow.
Here, we note that HLL scheme is better than 
simple Lax-Friedrich scheme 
in order to reproduce such a sharp discontinuity
\citep{2007A&A...464..429G}.
In contrast with M-1 method,
the shadow does not appear 
in the case of Eddington model.
As we have mentioned, 
since the isotropic radiation fields are assumed, 
the radiation comes around behind the clump
even without scattering.

Figure \ref{fig:shadow1} also shows that the radiation energy 
propagates with speed of light for M-1 model.
The dashed line in the left lower panel
indicates a wave front computed from $x=-0.5\ \mathrm{cm} + ct$.
The resulting wave front is in good agreement with
the dash line.
On the other hand, 
a wave speed reduces to $c/\sqrt{3}$ in the
Eddington model as we have discussed above. 
In the upper left panel,
we find that the position of a wave front 
is $x\sim 2.2$ km, which is consistent with
the estimation of $x=-0.5\ \mathrm{cm} + ct/\sqrt{3}$
with $t=41.8~\mathrm{\mu s}$.

We stress here again about the advantage of implicit 
treatment for gas-radiation interaction (source terms).
In this problem, the timescale of the gas-radiation interaction,
$(\rho_1 \kappa c)^{-1}$,
is around $1.1\times 10^{-10}\ \mathrm{s}$ in the clump,
which is much shorter than the timestep, $8.3\times 10^{-8} \mathrm{s}$.
If we explicitly integrate the gas-radiation interaction terms,
the numerical instability is caused.
We can take longer timestep via the implicit treatment.


\subsection{Tests for radiation hydrodynamics}\label{rad-HD}
In this subsection, we show the qualitative difference between the M-1 closure 
scheme and the Eddington approximation by solving the shock tube problems proposed
by \cite{2008PhRvD..78b4023F}. Although they obtained semi-analytic
solutions by assuming the Eddington approximation, there are no analytic
solutions with the M-1 closure due to the non-linearity in the Eddington
tensor. Since our numerical code can recover their analytical solutions
by adopting Eddington approximation in place of the M-1 closure
\citep{2013ApJ...764..122T}, we clearly understand the feature of 
the M-1 closure scheme and difference from the Eddington approximation.

A simulation box is bounded by $x=[-L,L]$, where $L=20$ in the
normalized unit. A number of grid points is fixed with $N_x=3200$ 
in this subsection.
Unlike \cite{2008PhRvD..78b4023F}, initially the
discontinuity is situated at $x=0$. 
The gas and radiation are in local thermal equilibrium in both sides
($x>0$ and $x<0$). 
The free boundary condition is applied in both boundaries
($x=-L$ and $x=L$). 
We again take the light speed as unity.
The Stefan-Boltzmann constant has a fictitious value of $a_R =
E'_{r,L}/T^4_{g,L}$, which is used to evaluate $E'_{r,R}=a_r T_{g,R}^4$
\citep{2008PhRvD..78b4023F,2011MNRAS.417.2899Z}, where the dash denotes a
quantity defined in the comoving frame, and subscripts $L$ and $R$
denote the left ($x<0$) and right ($x>0$) states, respectively. A
parameter set of initial conditions is summarized in
Table~\ref{tab:testparam}.

\subsubsection{non-relativistic strong shock}\label{nrs}
\begin{figure}
 \includegraphics[width=8cm]{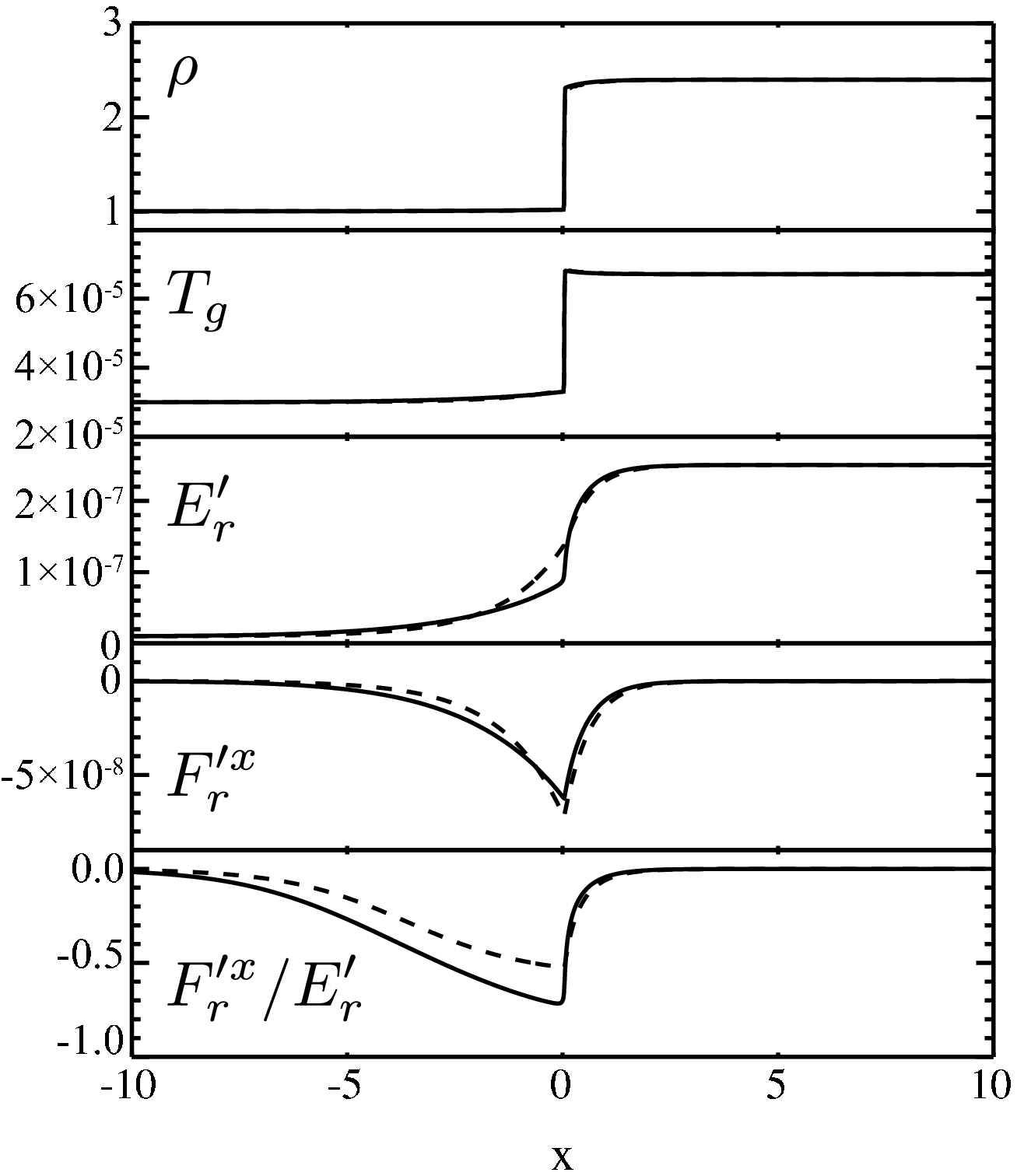}
  \caption{1-dimensional plots of the mass density, gas temperature,
 radiation energy density measured in the comoving frame, radiative flux
 measured in the comoving frame,
 and $F_r'^x/E_r'$} for the non-relativistic strong shock. Solid and
 dashed curves denote results with the M-1 closure and the Eddington
 approximation, respectively. 
  \label{fig:shtb1}
\end{figure}

Figure~\ref{fig:shtb1} shows the result of 
a non-relativistic strong shock problem (model RHDST1). 
We plot the mass density, gas temperature,
radiation energy density and radiation flux 
measured in the comoving frame, and the ratio of $F_r'$ to $E_r'$ at
$t=5000$ from top to bottom.
Solid curves denote solutions of M-1 model, while 
dashed curves represent numerical solutions of the Eddington
model.

Since the radiation energy density is much less than the gas
energy density, 
and since the radiation force does not play an important role,
the mass density and the gas temperature have 
a sharp discontinuity at the shock $(x=0)$ 
like a shock tube problem with pure hydrodynamics.
We also find that 
there is fewer differences between two models
as for the profiles of $\rho$ and $T_g$.
In contrast, the profiles of $E'_r$ and $F'^x_r$
by M-1 model are slightly different for those by Eddington model.
The radiation energy is transported from the shock front ($x=0$)
to the pre-shocked region ($x<0$) in both models.
The radiation flux is approximately given as $-E'_r$ 
for the M-1 model and as $-E'_r/\sqrt{3}$ for the Eddington model
(note that the speed of light is set to be unity in this subsection),
so that the ratio of $F'^x_r$ to $E'_r$ 
is smaller at $x\leq 0$ for the M-1 model than for the Eddington model
(bottom panel).
In addition, 
the radiation field is attenuated at the precursor region
via the absorption in both models.
However, we find that 
the gradient of the profiles of $E'_r$ and $F'^x_r$
are smoother in the M-1 model than in the Eddington model
(see the region of $x<0$).
The radiation field reduces 
with a distance from the shock front, 
$\propto e^{-\rho\kappa|x|}$, for the M-1 model,
but $\propto e^{-\sqrt{3}\rho\kappa|x|}$, for the Eddington model.
Such a difference is induced by that
the propagating speed of the radiation is decreased 
as $1/\sqrt{3}$ in the Eddington model
as we have discussed above.


\subsubsection{relativistic shock}\label{rs}
\begin{figure}
 \includegraphics[width=8cm]{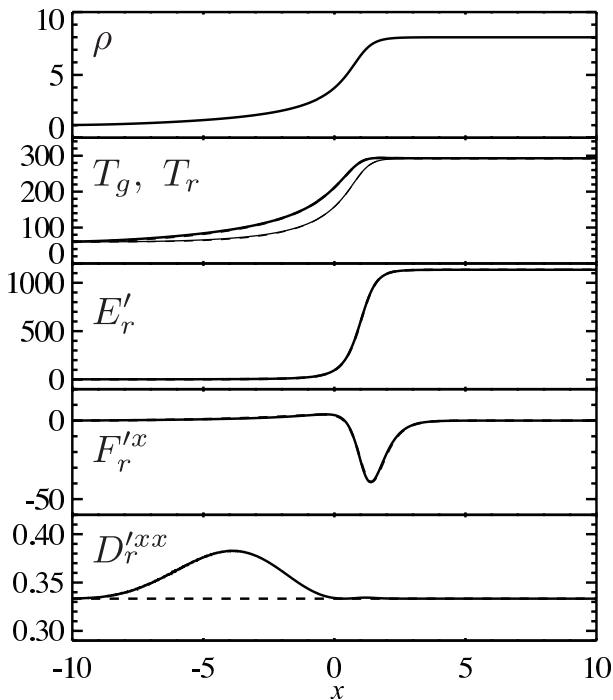}
  \caption{1-dimensional profiles of the relativistic shock. 
 From the top to bottom panels, the mass density, gas (thick curve) and
 radiation (thin curve) temperatures, radiation energy density,
 radiation flux, and
 $xx$-component of the Eddington tensor, are plotted, These quantities
 are measured in the comoving frame.
 Solid and dashed curves respectively shows the results with M-1 and
 Eddington models.}
  \label{fig:shtb2}
\end{figure}

A second shock tube problem is a relativistic shock including a radiation
(model RHDST2). 
Here, four velocity in the upstream is taken to be 10. 
Figure~\ref{fig:shtb2} 
shows profiles of mass density, temperatures of gas (thick curves) and
radiation (thin curves), radiation energy density, flux and
$xx$-component of the Eddington
tensor at $t=5000$ from top to bottom. 
Solid and dashed curves denote for solutions with the M-1 closure and
the Eddington approximation. 
In this test, 
the shock front is stationary for the Eddington model, 
but it very slowly moves with a speed of $1.6\times 10^{-4}$ 
for the M-1 model. 
In this figure,
the position of shock front is readjusted 
so as to be located at the origin
in order to compare solutions between two models.

We can see that solutions (except for $D'^{xx}$)
between two models are 
qualitatively and quantitatively consistent.
This is because that 
the optical depth is large enough,
and, then, the Eddington approximation is valid.
However, we find that 
$D_r'^{xx}$ slightly deviates from $1/3$ for the M-1 model,
although $D_r'^{xx}$ is $1/3$ by definition for the Eddington model.
The radiation energy is transported from the shock front
to the precursor region, leading to the slight 
anisotropic radiation field.
In our M-1 model, the maximum of $F'^x_r/E_r'$ is $\simeq 0.31$
and then we find $D_r'^{xx}=0.38$. 


Here, we note that the gas temperature is higher than the radiation
temperature in the preshocked region,
although
the gas temperature could not exceed the radiation temperature,
$T_g\leq T_r$,
if the gas is mainly heated up by absorption.
We confirmed that the compression is the dominant
heating mechanism.

\subsubsection{radiation pressure dominated shock}\label{rds}
In a radiation dominated mildly relativistic shock problem (model RHDST3),
the upstream radiation energy
density is set to be 20 times larger than the gas internal energy,
although the ratio is $2.2\times 10^{-4}$ and $3.3\times 10^{-2}$ for
tests in \S \ref{nrs} and \S \ref{rs}, respectively.
%

\begin{figure}
 \includegraphics[width=8cm]{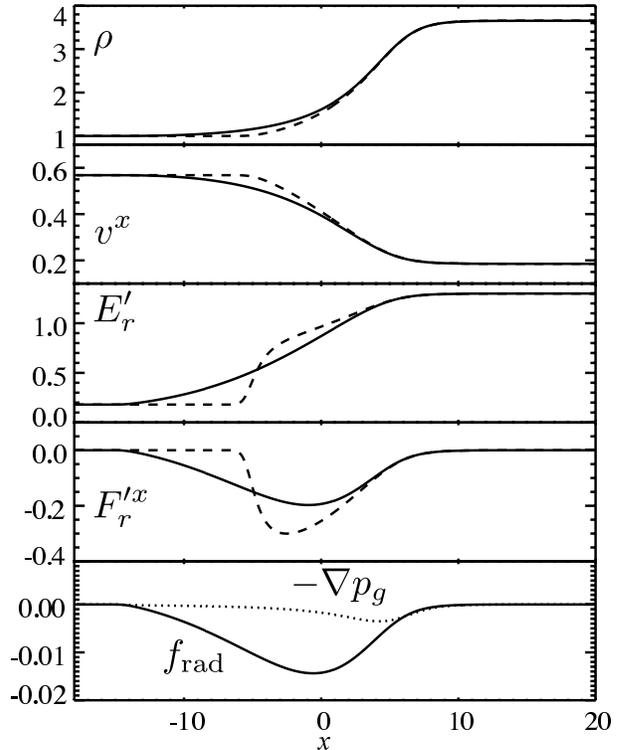}
  \caption{1-dimensional plot of the mass density, $v^x$,
 radiation energy density measured in the comoving frame, radiation flux
 measured in the comoving frame, 
 and forces acting on the plasma for the radiation shock. In the top
 four panels, solid and
 dashed curves denote the results with the M-1 closure and the Eddington
 approximation, respectively. In the bottom panel, the solid and dotted
 curves denote the radiative force and the gas pressure gradient force. }
  \label{fig:shtb3}
\end{figure}

Figure~\ref{fig:shtb3} shows profiles of $\rho, v^x, E'_r,
F'^x_r$, and forces acting on a gas at $t=5000$ from top to bottom. In
upper four panels, solid and dashed curves denote solutions of M-1
and Eddington models. In the bottom panel, solid and dotted curves
represent the radiation gas pressure gradient force for the M-1 model,
respectively. 
After an initial discontinuity at $x=0$ breaks up, 
a steady state solution,
in which the shock front is located at the origin,
gradually forms in the Eddington model.
In the case of the M-1 model, 
although the shock moves with a constant velocity of 
$v_{sh}=-5.3\times 10^{-4}$,
the profiles approach to a steady solution 
for the frame of reference in which the shock front is stationary.
Similar to two tests have shown in \S \ref{nrs} and \S \ref{rs}, 
a precursor wave propagates in a upstream region. 

It is found that the radiation flux is negative 
in both models (forth panel),
implying that the radiation energy is transported from right to left.
The leftward radiation flux ($-F'^x_r$), 
which is at maximum at around $x=0$ (M-1) or $x=-3$ (Eddington),
is reduced by absorption and approaches to null with decreasing $x$.
Although the radiation energy is transported 
up to merely $x\sim -7$ for Eddington model,
the leftward radiation penetrates to $x\sim -12$ for M-1 model.
Since the speed of light is effectively reduced to $1/\sqrt{3}$
as we have discussed above,
the leftward radiation flux suddenly decreases 
via the enhanced absorption for the Eddington model.
Therefore, in the M-1 model,
the profile of the radiation energy density is smooth,
the radiation energy density is enhanced 
even at the range of $x<-7$.

The leftward radiation flux in the upstream region
induces the leftward radiation force (bottom panel),
which works to decrease the velocity.
Hence, the velocity (density) starts to decrease (increase)
at $x\sim -12$ for M-1 model
and $x\sim -7$ for Eddington model
(see top and second panels).
The bottom panel clearly shows that 
the pressure gradient force is weaker than the radiation force.
Here we note that,
in contrast with RHDST1,
the precursor strongly affects on the upstream gas
since a radiation energy much exceeds a gas energy density
in the present test.



\begin{figure*}
 \begin{center}
 \includegraphics[width=16cm]{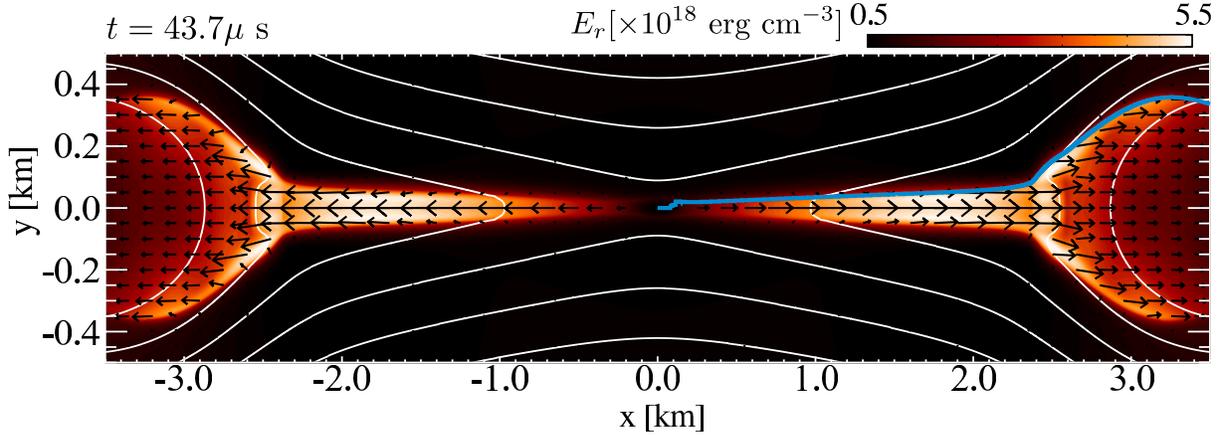}
  \caption{Color shows $E_r$ at $t=47.3\ \mu\mathrm{s}$, while white
 curves and arrows denote for  magnetic field lines and radiation flux, respectively.
 A blue line denotes a position at which there is a steep jump on $j_z$
 in the first quadrant.}
  \label{fig:mrxcol}
 \end{center}
\end{figure*}

\subsection{Relativistic Petschek Type Magnetic Reconnection with Radiation Field}\label{r3recon}
Lastly, we perform a SR-R2MHD simulation of a relativistic magnetic
reconnection. Recently, some authors have studied the relativistic
magnetic reconnection without radiation by assuming uniform resistivity
model 
\citep{2011ApJ...739L..53T} and a spatially localized resistivity
model \citep{2006ApJ...647L.123W, 2010ApJ...716L.214Z,
2011MNRAS.418.1004Z}. 
Also the importance of the radiative effects on
the magnetic reconnection is studied \citep{1984ApJ...276..391S,
2009PhRvL.103g5002J, 2011PhPl...18d2105U}.
In this section, 
we adopt a spatially localized resistivity model for
a fast (Petschek type) magnetic reconnection.

We solve SR-R2MHD equations in the Cartesian coordinate on the $x-y$
plane. 
A computational domain consist of $x=[0,\ 17.4]\
\mathrm{km}$ and $y=[0,\ 5.7]\ \mathrm{km}$. We use non uniform grids
and a number of grid points is $(N_x, N_y)=(3500,\ 800)$. A
minimum grid size is $\Delta x= \Delta y=100\mathrm{cm}$. 
Because of the symmetry of the system, we adopt a point symmetric
boundary condition. Scalar quantities are symmetric at $x=0$ and $y=0$.
At $x=0$, $u^y, B^x, B^z, E^x, E^z, F_r^y$ are symmetric while the rest
of the vector components are anti-symmetric.
At $y=0$, $u^x, B^y, B^z, E^y, E^z, F_r^x$ are symmetric and the vector
components are anti-symmetric. 
The free boundary conditions are applied at the other boundaries. 
We assume an isothermal and uniform gas,
$\rho_0=0.01~\mathrm{g~cm^{-3}}$ and $T_g=10^8~\mathrm{K}$ in a whole
domain at the initial state. 
The gas is initially in LTE, $T_g=T_r$.
We assume a force free magnetic field configuration given by
\begin{eqnarray}
 \bmath B = B_0 \tanh\left(\frac{y}{\lambda}\right)\bmath e_x 
  + B_0 \mathrm{sech}\left(\frac{y}{\lambda}\right)\bmath e_y,
\end{eqnarray}
\citep{1973ApJ...181..209L, 2007MNRAS.374..415K}, 
where $B_0=10^{10}~\mathrm{G}$ is an amplitude of the magnetic field and
$\bmath e_x$ and $\bmath e_y$ are unit vectors in $x$- and
$y$-direction. $\lambda=10^4~\mathrm{cm}$ is a thickness of a
current sheet. 
Since we take a mean molecular weight to be $0.5$,
the plasma-$\beta$ in initial state is $4.1\times 10^{-5}$.

We adopt a spatially localized resistivity model to attain the fast
magnetic reconnection:
\begin{equation}
 \eta = \eta_u + \frac{\eta_i - \eta_u}{\cosh[(x^2+y^2)/\lambda]^2},
\end{equation}
where $\eta_u$ and $\eta_i$ are constants.
We set corresponding magnetic Reynolds numbers as
$R_{M,u}=4\pi \lambda c/\eta_u=400$ and $R_{M,i}=4\pi \lambda
c/\eta_i=50$. 
For opacity, we assume electron scattering and free-free absorption.
The typical optical depth is $40$ for scattering 
and $6.4\times 10^{-6}$ for absorption.

Figure\ \ref{fig:mrxcol} shows results at $t=47.3\mu\mathrm{s}$. 
Color, arrows, and white curves indicate the radiation energy
density, flux, and magnetic field lines in the observer frame. 
A blue line denotes a position at which there is a steep jump on $j_z$
in the first quadrant.

Due to an enhancement of the electric resistivity at the origin,
magnetic field lines start to reconnect and the gas is evacuated as
outflows in the $\pm x$-directions. 
Since we 
adopt a spatially localized resistivity model, four slow shocks attached
to the diffusion region form (one of them is indicated by a blue
curve) \citep{2006ApJ...647L.123W, 2010ApJ...716L.214Z, 2011MNRAS.418.1004Z}. 
This indicates that a fast Petschek type magnetic reconnection 
is realized even though the radiation field is fulfilled. 

We can see that the radiation energy density is confined 
in exhausts of outflows.
The photons suffer from numerous scattering,
since the system is very optically thick for scattering.
Thus, the radiation flows together with matter via advection,
and $\bmath F_r$ is almost parallel to velocity fields. 
This implies that 
the reconnection region is very brightly observed 
for downstream observers (on $x$-axis).
On the other hand, it would be difficult to detect the reconnection region 
for observers around $y$-axis or $z$-axis.

\begin{figure}
 \includegraphics[width=8cm]{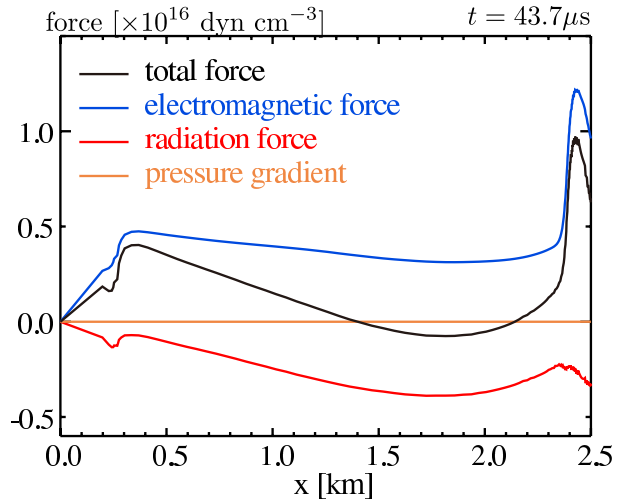}
  \caption{Total (black), electromagnetic (blue), radiation
 (red) and pressure gradient (orange) force densities acting on the fluids
 along the slow shock.}
\label{fig:mrxforce}
\end{figure}

In order to consider radiation effects on the dynamics,
we plot in Figure \ref{fig:mrxforce} 
the pressure gradient force including effects of enthalpy variation
(blue),
electromagnetic force with non-adiabatic term (red), 
radiation force (orange), 
and total force density (black) 
along a slow shock denoted by a blue line in Figure \ref{fig:mrxcol}. 
This figure clearly shows that 
the electromagnetic force accelerates the gas
and the pressure gradient force is negligible.
The matter is decelerated by the radiation force.
Such a deceleration is caused by the radiation drag
($\propto (v^iE_r + v_j P_r^{ij}$)
which becomes non-negligible compared with the radiation flux force
($\propto F_r$), when $F^i_r<(v^iE_r+v_jP_r^{ij})$.
In the present problem, 
the condition of $F^i_r<(v^iE_r+v_jP_r^{ij})$ is moderately realised
since the large optical thickness reduces the radiation flux.
The typical value of $F^x_r/(v^xE_r+v_jP_r^{xj})$ is around unity.

In the present problem, 
the radiation drag is also non-negligible compared with 
the electromagnetic force.
Here, we recover a light speed $c$ to avoid misunderstanding.
The ratio of the radiation drag 
to the electromagnetic force is
$\rho \sigma_s v E_r l / 4\pi c \bmath B^2$,
where we assume a mildly relativistic plasma
and estimate the radiation drag and 
the electromagnetic force as 
$\sim \rho \sigma_s v E_r /c$ 
and $\sim 4\pi \bmath B^2/l$
with $l$ being a typical length of the current sheet
for the order estimation.
Such a ratio is rewritten as
$0.5 \tau_\mathrm{cs} (E_r /E_{\rm mag}) (v/c)$,
where $E_\mathrm{mag}(=\bmath B^2/8\pi)$ 
is the magnetic energy density and
$\tau_\mathrm{cs}(=\rho l \sigma_s)$ is the optical depth 
of the current sheet.
It implies that 
the radiation drag tends to play an important role
for magnetic reconnection 
in the high density and high velocity plasma.
In our simulations,
we have $\tau_\mathrm{cs} (E_r /E_{\rm mag}) (v/c) \sim 3.6$ by assuming
$l = \lambda$ and $v = v_A=B/\sqrt{4\pi \rho}$.
Due to the radiation drag,
the outflow four
velocity is about 10\% slower with the radiation field than without
the radiation field in our parameter set.

\begin{figure}
 \includegraphics[width=8cm]{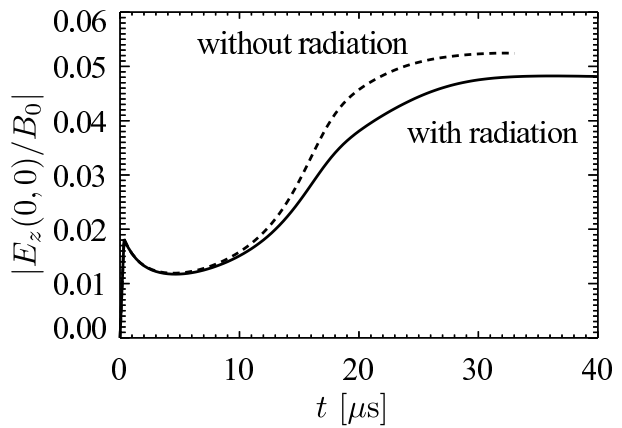}
  \caption{Time evolution of the reconnection rate $E_z(0,0)/B_0$. Solid
 and dashed curves denote for results with and without radiation fields,
 respectively.}
\label{fig:mrxrate}
\end{figure}

In Figure~\ref{fig:mrxrate}, we show a time evolution of
reconnection rate, which is here defined by $E_z(0,0)/B_0$
($x$- and $y$-components of the electric fields are null by definition).
Solid and dashed curves are results with and without
solving radiation field, respectively. 
Due to an enhancement of the localized resistivity, 
magnetic field lines start to reconnect and amplitude of electric field
rapidly increases by dissipating the magnetic energy. 
After $t=30\ \mu\mathrm{s}$, 
quasi steady state is realized 
and the reconnection rate roughly becomes constant.
The reconnection rate at the steady state
is about 10\% smaller with the radiation field than
without the radiation field. 
It is understood as below.
As we have already mentioned, 
the radiation drag force slows down the outflow velocity. 
Then, the inflow velocity in the quasi steady state
(downward and upward component of the velocity 
in the regions of $y>0$ and $y<0$) 
is also reduced.
Thus, the $z$-component of the electric fields,
[$E_z = -(\bmath v\times \bmath B)_z/c$], decreases,
inducing the reduction of the magnetic reconnection rate.
%

Although we show the results for one parameter set,
the reconnection rate as well as the outflow velocity,
would depend on initial parameters.
The systematic study of the magnetic reconnection with radiation fields
will be reported in the forthcoming paper. 

\section{Summary}\label{summary}
We developed a special relativistic radiation-magnetohydrodynamic (SR-RMHD) code,
in which the M-1 closure method is employed
and a source term for gas-radiation interaction
is implicitly and iteratively integrated.
We also extend our SR-RMHD code to a special
relativistic resistive radiation-MHD (SR-R2MHD) code, 
which includes electric resistivity.
Our SR-RMHD code successfully solves some of test problems,
i.e., shock tube problems of MHD/Radiation-HD and
propagating radiation,
and we demonstrate the radiation drag effect
in relativistic Petschek type magnetic reconnection
by SR-R2MHD code.

Since our code use radiation fields only in the observer's frame,
we straightforwardly compute $\bmath P_r$ from $E_r$ and $\bmath F_r$
through the M-1 closure method without the Lorentz transformation.
In contrast, the Lorentz transformation is inevitable
for the Eddington approximation.
By virtue of M-1 closure method,
anisotropic propagation of radiation is solved 
and the propagating speed of the radiation is $c$ 
in the optically thin media.
The Eddington approximation as well as 
flux-limited diffusion approximation
is problematic for such anisotropy.
In addition, the speed of light is reduced 
to be $c/\sqrt{3}$ for the Eddington approximation.

In our code, 
all of advection terms are explicitly integrated
by setting the timestep to be a fraction of 
$\Delta x/c$ with $\Delta x$ being the grid spacing.
Implicit integration of the source term prevents
the timestep from shortening
when the timescale of the source term 
(e.g., gas-radiation interaction) becomes very small.
For the implicit treatment,
we directly invert a $4\times 4$ matrix at each grid point in SR-RMHD code.
In addition to the gas-radiation interaction term,
the source term appeared in Ampere's law
is also solved implicitly in SR-R2MHD code.
Then, we need to invert $3\times 3$ and $4\times 4$ matrices 
at each grid point.
Such matrix inversion is carried out analytically
without communication with neighbor grids.
Thus, our code could be massively parallelized without difficulty.
Our code would be widely utilized for the relativistic 
astrophysical phenomena,
even though the dense and less dense regions are mixed.

\acknowledgments
We are grateful to an anonymous referee for improving our manuscript.
Numerical computations were carried out on Cray XT4 at the Center for
Computational Astrophysics, CfCA, at the National Astronomical Observatory
of Japan, on Fujitsu FX-1 at the JAXA Supercomputer System (JSS) at the Japan
Aerospace Exploration Agency (JAXA), and on T2K at the University of
Tokyo. This work is supported in part by Ministry of Education, Culture,
Sports, Science, and Technology (MEXT) for Research Activity Start-up
23840045 (HRT) and Young Scientist (B) 24740127 (K.O.). 
A part of this research has been funded by MEXT HPCI
STRATEGIC PROGRAM.


\end{document}